\pdfoutput=1
\documentclass[11pt]{article}
\usepackage{acl}

\usepackage{times}
\usepackage{latexsym}

\usepackage[T1]{fontenc}
\usepackage[utf8]{inputenc}

\usepackage{microtype}
\usepackage{inconsolata}

\usepackage{graphicx}
\usepackage{amsmath}
\usepackage{amssymb}
\usepackage{pifont}
\usepackage{tabularx}
\usepackage[table]{xcolor}
\usepackage{booktabs}
\usepackage{multirow}
\usepackage{amsthm}
\usepackage{comment}

\usepackage{enumitem}

\definecolor{lightblue}{RGB}{235,242,255}
\definecolor{lightgray}{RGB}{240,240,240}

\usepackage[most]{tcolorbox}
\newtcolorbox{promptbox}{
  enhanced,
  breakable,
  colback=gray!5,
  colframe=gray!60,
  boxrule=0.5pt,
  arc=1mm,
  boxsep=2pt,
  left=4pt,
  right=4pt,
  top=4pt,
  bottom=4pt,
  before skip=4pt,
  after skip=6pt,
  fontupper=\small,
  before upper=\raggedright,
  before lower=\raggedright
}

\usepackage{balance}

\title{HyperSU: Corpus-Driven Semantic-Unit Hypergraph for Retrieval-Augmented Generation}

\author{
\textbf{Jiate Liu}\textsuperscript{1}\thanks{\ Equal contribution.} \quad
\textbf{Liuyi Chen}\textsuperscript{2}\footnotemark[1] \quad
\textbf{Zhengyi Yang}\textsuperscript{3}\thanks{\ Corresponding author.} \quad
\textbf{Chuan He}\textsuperscript{1} \\
\textbf{Mingchen Ju}\textsuperscript{4} \quad
\textbf{Bocheng Han}\textsuperscript{5} \quad
\textbf{Ruyi Liu}\textsuperscript{1} \quad
\textbf{Xu Zhou}\textsuperscript{2} \\[2pt]
\textsuperscript{1}The University of New South Wales, Sydney, Australia \quad
\textsuperscript{2}Hunan University, Changsha, China \\
\textsuperscript{3}The University of Sydney, Sydney, Australia \quad
\textsuperscript{4}Euler AI \quad
\textsuperscript{5}Vecton AI \\[2pt]
\texttt{\{jiate.liu, chuan.he3, ruyi.liu2\}@unsw.edu.au} \quad
\texttt{\{liuyi.chen, zhxu\}@hnu.edu.cn} \\
\texttt{zhengyi.yang@sydney.edu.au} \quad
\texttt{mingchen.ju@eulerai.au} \quad
\texttt{bocheng.han@vectonai.com}
}

\begin{document}
\maketitle

\begin{abstract}
Recent Hypergraph-based retrieval-augmented generation (HyperRAG) methods use hyperedges to connect multiple entities simultaneously, enabling more efficient multi-entity evidence organization than pairwise graph structures.  However, existing HyperRAG methods often rely on LLM-generated summaries to construct hyperedges, which can introduce hallucinations while also incurring high indexing costs. In addition, during retrieval, existing methods typically rely on either one-hop neighbor expansion or PageRank diffusion. The former may miss useful multi-hop evidence, while the latter can suffer from uncontrolled propagation over excessive hub nodes, leading to semantic drift and noisy reasoning chains.
To address these challenges, we propose \textbf{HyperSU}, a novel hypergraph-based RAG framework featuring semantic-unit hyperedges and clue-guided bidirectional retrieval. During construction, \textbf{HyperSU} formulates hyperedge construction as an entity-aware minimum-description-length (MDL) optimization problem, inducing source-grounded semantic-unit hyperedges that balance sentence-level semantic coherence and entity compactness. It then constructs a hypergraph by modeling each semantic unit as a hyperedge over its co-mentioned entities. During retrieval, HyperSU performs clue-guided bidirectional expansion over the semantic-unit hypergraph, enabling both multi-hop evidence discovery and evidence-anchored noise reduction. Experiments show that \textbf{HyperSU} consistently improves answer accuracy over standard, graph-based, and hypergraph-based RAG baselines, raising average accuracy on GraphRAG-Bench by 5.5 points over the strongest baseline (68.7 vs.\ 63.2), with larger gains on reasoning-intensive tasks, where the relative improvement reaches 14.1\%.
\end{abstract}

\section{Introduction}

Retrieval-augmented generation (RAG) enhances large language models by grounding generation on retrieved external knowledge. ~\citep{lewis2020retrieval,gao2024rag_survey}. 
However, traditional retrieval methods often struggle to capture complex multi-hop relationships across dispersed evidence. To address this limitation, recent graph-based RAG methods~\citep{edge2024graphrag,guo-etal-2025-lightrag,gutierrez2024hipporag} organize corpus knowledge as graphs to support multi-hop retrieval and reasoning.
Since each edge connects only two entities, graph-based methods are limited to binary relations and struggle to model common multi-entity relations in real-world knowledge.  Hypergraph-based RAG methods address this limitation by using hyperedges to connect multiple related entities simultaneously. 

\begin{figure*}[t]
    \centering
    \vspace{-0.2cm}
    \includegraphics[width=\textwidth,trim=3 3 3 3,clip]{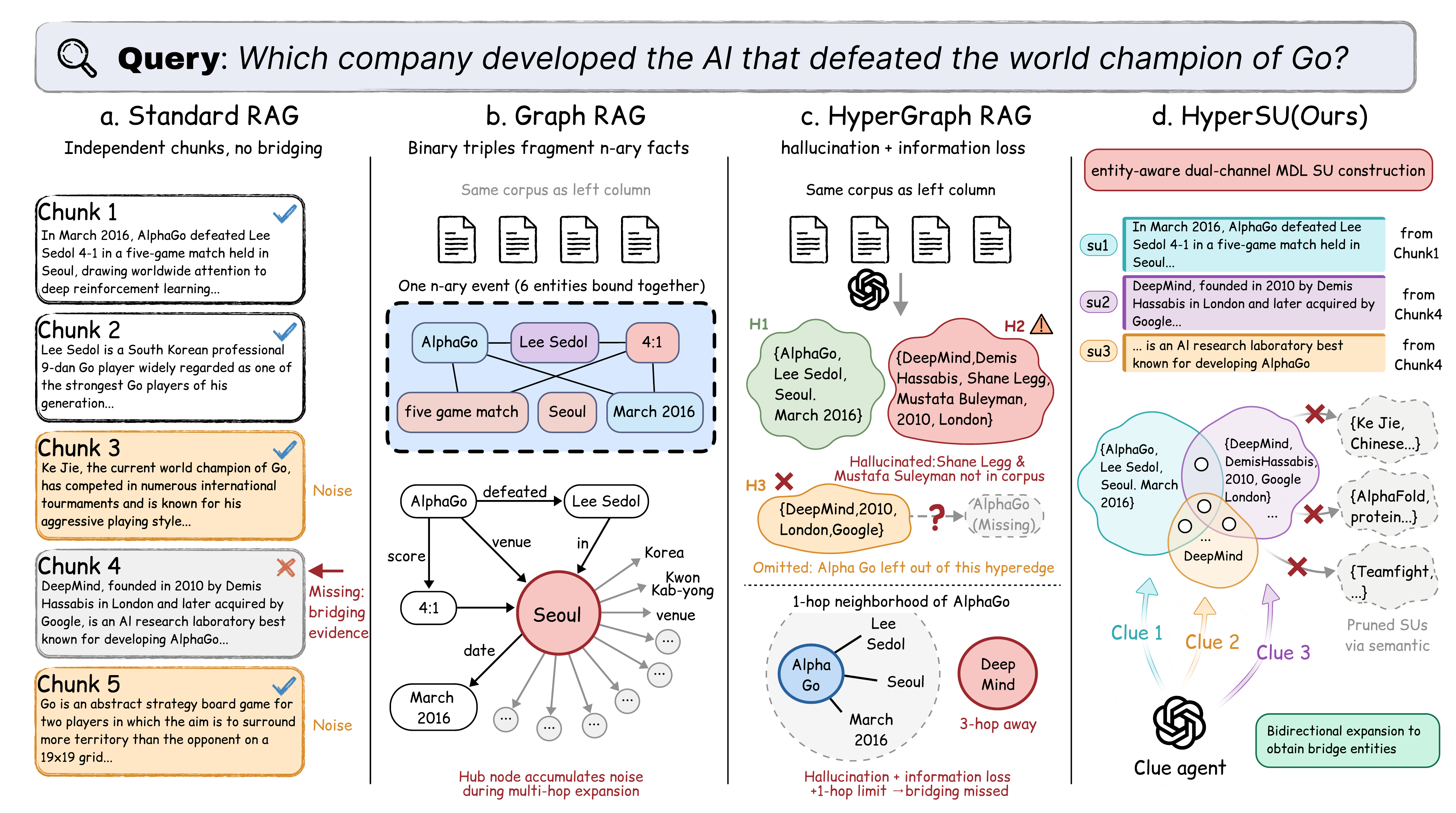}
    \caption{Comparison of RAG paradigms on a multi-hop QA example. Unlike Standard RAG, GraphRAG, and existing hypergraph-based RAG, HyperSU builds source-grounded semantic-unit hyperedges and retrieves bridging evidence through clue-guided bidirectional expansion.}
    \vspace{-0.2cm}
    \label{fig:comparison}
\end{figure*}

As illustrated in Figure~\ref{fig:comparison}(a)-(b), consider the query: \textit{``Which company developed the AI that defeated the world champion of Go?''}

RAG retrieves independent chunks without bridging evidence, introducing noisy context (\textit{Ke Jie} and \textit{Go}) while missing critical connecting evidence such as \textit{DeepMind}.
GraphRAG improves 
retrieval by organizing knowledge as relational
graphs. However, each edge connects only 
two entities, a multi-entity relation must be decomposed into multiple binary triples linked through 
shared hub entities such as \textit{Seoul}, fragmenting a 
coherent n-ary relation into multiple pairwise
connections.

Existing HyperRAG methods ~\citep{luo2025hypergraphrag,feng2026hyperrag}, however, commonly rely on LLM-generated hyperedges and often retrieve through one-hop neighboring expansion or PageRank~\citep{page1999pagerank} diffusion. 
This design still has two limitations. 
First, generated hyperedges may contain unsupported or incomplete relations.
For example, in Figure~\ref{fig:comparison}(c), hyperedge $H_2$ incorrectly includes \textit{Shane Legg} and \textit{Mustafa Suleyman}, while $H_3$ omits the key entity \textit{AlphaGo}. 
Second, one-hop expansion can miss distant but necessary bridge evidence, while unrestricted PageRank diffusion may introduce excessive noisy evidence and semantic drift.
In this example, \textit{DeepMind} remains outside the one-hop neighborhood of \textit{AlphaGo}, leaving the reasoning chain incomplete.

These limitations motivate us to develop a 
framework that constructs hyperedges that \textbf{reduce dependence on LLM-generated summaries}, while supporting \textbf{effective retrieval} for more complete and reliable answers.

To achieve this goal, we propose a novel framework, \textbf{HyperSU}. During construction, HyperSU formulates semantic-unit construction as a minimum description length (MDL) optimization problem~\citep{rissanen1978mdl,grunwald2007mdl} that jointly optimizes sentence-level semantic coherence and entity compactness.
During retrieval, HyperSU decomposes the query into multiple clues and performs clue-guided bidirectional expansion over the semantic-unit hypergraph. Forward expansion discovers distant bridge entities to improve multi-hop recall, while backward anchoring verifies forward-reached evidence against dense-retrieved candidates to suppress semantic drift. The activated SUs are finally mapped back to their source passages for ranking and answer generation.

HyperSU constructs a semantically coherent and entity-compact hypergraph that reduces redundant and noisy evidence. Meanwhile, its retrieval strategy improves evidence coverage while controlling irrelevant propagation.
As shown in Figure \ref{fig:comparison}(d), each semantic unit (SU) captures a coherent event or factual context. For example, the hyperedge marked in blue, \{\textit{AlphaGo}, \textit{Lee Sedol}, \textit{Seoul}, \textit{March 2016}\}, describes the \textit{AlphaGo}--\textit{Lee Sedol} match held in \textit{Seoul} in \textit{March 2016}. Shared entities marked in white naturally connect related SUs, forming an entity-compact hypergraph with strong semantic associations across dispersed evidence. During retrieval, query-based clues guide bidirectional exploration over the hypergraph, while irrelevant SUs such as \textit{Ke Jie}, \textit{AlphaFold} and \textit{Teamfight} are effectively pruned to suppress noisy propagation.

Our main contributions are as follows:

\begin{itemize}[topsep=1pt,itemsep=1pt,parsep=0pt,partopsep=0pt,leftmargin=*]
    \item  We propose \textbf{HyperSU}, a corpus-driven HyperRAG framework built on semantic units. HyperSU constructs a semantically coherent hypergraph with compact entity connections, where hyperedges are grounded in complete and faithful corpus evidence.
    \item We design a clue-guided bidirectional hypergraph retrieval algorithm that combines fault-tolerant multi-clue SU activation, recall-oriented forward expansion, precision-oriented backward anchoring, and convergence verification, enabling reliable bridge-evidence discovery while suppressing semantic drift in multi-hop reasoning.
    \item Experiments on GraphRAG-Bench, HotpotQA, 2WikiMultiHopQA, and MuSiQue show that HyperSU consistently improves over standard, graph-based, and hypergraph-based RAG baselines, raising average GraphRAG-Bench accuracy by 5.5 points over the strongest baseline (68.7 vs.\ 63.2), with a maximum relative gain of 14.1\% on a single setting, and obtaining the best GPT-Accuracy across three multi-hop QA datasets. HyperSU further constructs hyperedges with zero offline LLM indexing cost, requiring 0 indexing tokens and \$0 API cost.\footnote{Throughout the paper, zero LLM indexing cost means zero offline generative-LLM API calls and tokens. Indexing still runs GLiNER, an encoder-based NER model, locally, and its wall-clock cost is reported in \S\ref{sec:efficiency}; an LLM is used online by the Clue Agent, whose per-query cost is also reported there.}
\end{itemize}

\section{Related Work}

\noindent\textbf{Graph-based RAG}
Graph-based RAG methods augment retrieval by organizing corpus evidence into structured representations.
Existing systems differ in how they build and traverse such structures: community-summary methods construct graph communities and summaries~\citep{edge2024graphrag,lazygraphrag2024}; PageRank- or traversal-based methods expand from query-linked entities~\citep{guo-etal-2025-lightrag,fastgraphrag2024,gutierrez2024hipporag,gutierrez2025hipporag2}; learned, prompted, or agentic retrievers use neural or LLM-guided mechanisms to select relevant subgraphs~\citep{luo2025gfmrag,wang2024kgp,he2024gretriever,liu2026a2rag}; and hierarchical, relation-free, or hybrid methods organize evidence beyond standard binary KG retrieval~\citep{sarthi2024raptor,linearrag2026,zhao2025e2graphrag}.
While these methods improve structured retrieval, binary or summary-based representations may fragment multi-entity evidence or require costly LLM-based indexing.
A detailed comparison is provided in Appendix~\ref{app:method-comparison}.

\noindent\textbf{Hypergraph-based RAG}
Hypergraphs are widely used to represent higher-order relations in graph learning and analytics~\citep{luo2024hierarchical,li2026fairness,pan2025hitec}.
Hypergraph-based RAG methods model higher-order evidence with hyperedges rather than pairwise graph edges.
HyperGraphRAG~\citep{luo2025hypergraphrag} constructs local n-ary relational hyperedges from pre-segmented chunks, while Hyper-RAG~\citep{feng2026hyperrag} models higher-order correlations through hypergraph-driven retrieval.
However, these hyperedges or correlations are typically generated from chunk-level text by LLMs, which can introduce unsupported entities, omit source-critical bridge entities, and increase offline indexing cost.
HyperSU instead induces source-grounded semantic-unit hyperedges from corpus spans and composes them at query time through clue-guided bidirectional expansion.
Because each hyperedge is a source span, retrieved evidence remains traceable to the original text, a property recent work argues data systems should provide by design~\citep{yang2026raids}.

\section{Preliminaries}
\label{sec:preliminaries}

\noindent\textbf{Definition 1: Retrieval-Augmented Generation.}
Given a question $q$ and a passage corpus $\mathcal{C}$, retrieval-augmented generation (RAG) retrieves a subset of passages $\mathcal{P}^* \subseteq \mathcal{C}$ and generates an answer $y$ conditioned on $(q,\mathcal{P}^*)$:
\[
  P(y \mid q) = \sum_{\mathcal{P}^* \subseteq \mathcal{C}}
  P(y \mid q, \mathcal{P}^*)\, P(\mathcal{P}^* \mid q, \mathcal{C}).
\]

\noindent\textbf{Definition 2: Hypergraph.}
A hypergraph $G_H = (V, E_H)$ generalizes a graph by allowing each hyperedge $e_H \in E_H$ to connect a set of entities
$V_{e_H} = \{v_1, v_2, \dots, v_n\}$, where $n \geq 2$.
It can be represented by an incidence matrix
$H \in \{0,1\}^{|V| \times |E_H|}$, where
$H[v, e_H] = 1$ iff $v \in V_{e_H}$.
Through star expansion~\citep{agarwal2006higherorder,zhou2006hypergraph,sun2008hypergraph},
$G_H$ can be equivalently encoded as an entity--hyperedge bipartite graph whose
biadjacency matrix is exactly $H$; we detail this correspondence in
Appendix~\ref{app:star_expansion}.

We use this formalism to represent evidence units that involve more than two co-mentioned entities. Unlike the pairwise entity--entity projection, a hyperedge retains the joint membership of all entities associated with the same source-grounded evidence unit.

\noindent\textbf{Proposition 1}
\textit{For a source-grounded multi-entity evidence unit $e$ with entity set $V_e$, representing $e$ as a hyperedge preserves the unit-level membership and provenance of its co-mentioned entities. In contrast, projecting $e$ onto pairwise entity--entity edges may lose the original group identity unless auxiliary event or provenance nodes or labels are added.}

We provide the formal proof in Appendix~\ref{app:proof_prop1}.

\section{Method}

\begin{figure*}[t]
    \centering
    \vspace{-0.45em}
    \includegraphics[width=\textwidth]{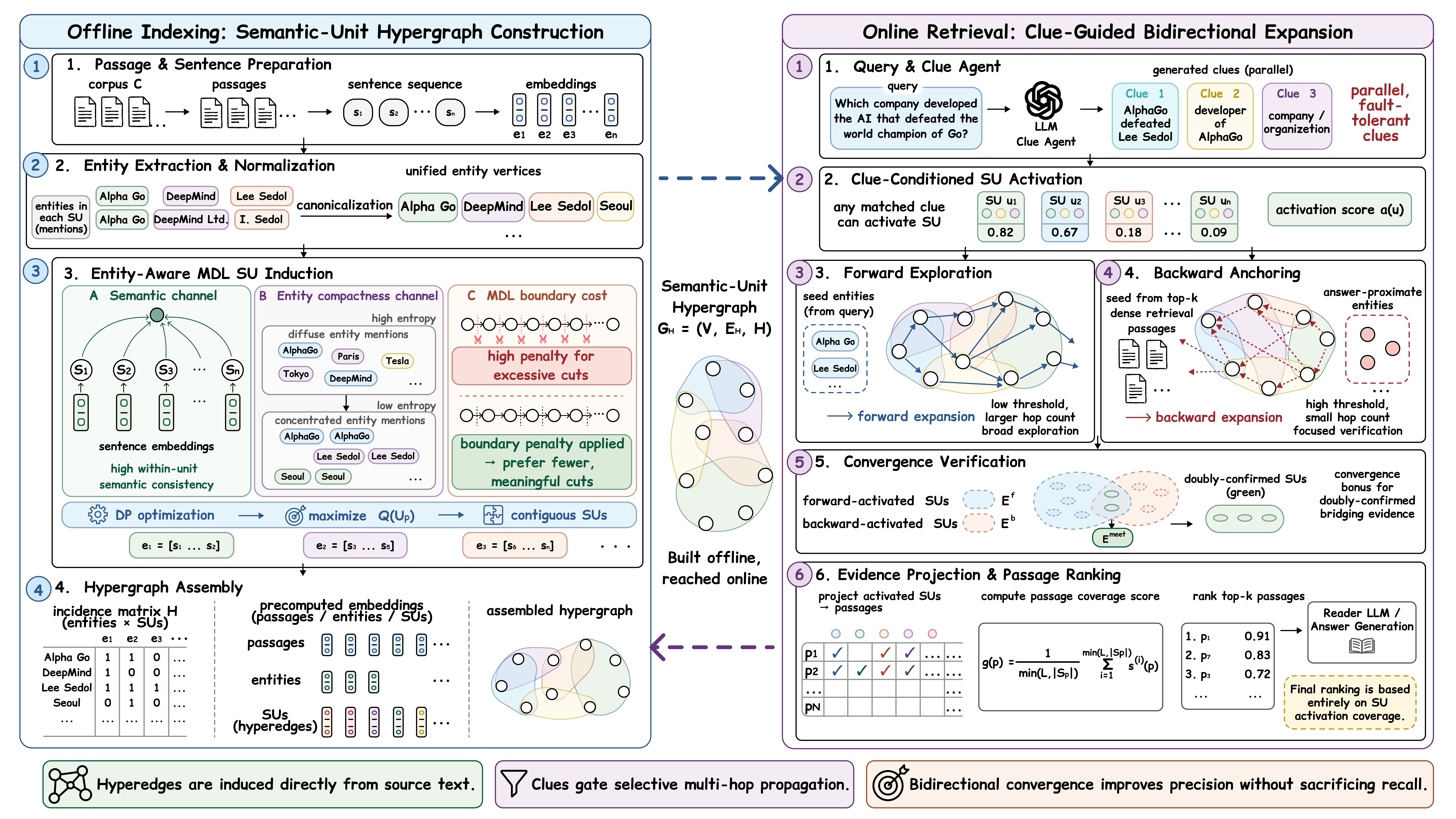}
    \vspace{-0.45cm}
    \caption{Overview of HyperSU. Offline indexing builds a source-grounded entity--SU hypergraph, while online retrieval activates, expands, and ranks SUs through clue-guided bidirectional retrieval.}
    \label{fig:overview}
    \vspace{-0.75em}
\end{figure*}

HyperSU consists of two stages: offline semantic-unit hypergraph construction and online clue-guided retrieval. As shown in Figure~\ref{fig:overview}, the offline stage first extracts sentence-level entity mentions, induces compact semantic units (SUs) through entity-aware MDL segmentation, and assembles the semantic-unit hypergraph $G_H=(V,E_H,H)$. The online stage composes these local evidence units into query-specific reasoning chains: it maps query entities to the hypergraph, derives parallel clues for SU activation, performs bidirectional frontier expansion, and verifies convergent evidence before projecting activated SUs back to passages for ranking and answer generation. We describe offline construction in Section~\ref{sec:indexing} and online retrieval in Section~\ref{sec:retrieval}.

\subsection{Offline Indexing: Semantic-Unit Hypergraph Construction}
\label{sec:indexing}

Given a corpus $\mathcal{C}$, HyperSU constructs the hypergraph through sentence-level entity extraction, entity-aware SU induction, and assembly.

\phantomsection
\label{sec:entity_extraction}
\noindent\textbf{Entity Extraction.}
For each passage $p=(s_1,\dots,s_n)$, HyperSU first performs sentence-level entity extraction using GLiNER~\citep{zaratiana2024gliner}, a generalist named entity recognition model. The extracted mentions are normalized into canonical entity names and used to construct the entity-count channel for SU induction.

\noindent\textbf{Definition 3: Semantic Unit.}
Given a passage $p=(s_1, s_2, \dots, s_n)$, a \emph{semantic unit} (SU) is a contiguous, semantically coherent, self-contained evidence span $u=(s_i,\dots,s_j)$. HyperSU represents each SU as a source-grounded local evidence unit and induces a hyperedge $e_u \in E_H$ over its co-mentioned entities, with $V_u=\{v\in V: v \text{ appears in } u\}$. SU hyperedges are reused and composed during retrieval to form query-specific reasoning chains, rather than pre-materialized as predicate-level n-ary facts. The construction is formalized as an MDL segmentation problem in \S\ref{sec:su_construction}.

\phantomsection
\label{sec:su_construction}
\noindent\textbf{Entity-Aware SU Induction.}
Let $(\mathbf{x}_1,\dots,\mathbf{x}_n)$ denote L2-normalized sentence embeddings. For each passage $p$, HyperSU partitions sentences into contiguous, non-overlapping semantic units $\mathcal{U}_p={u_1,\dots,u_k}$. We formulate this segmentation as an entity-aware dual-channel MDL problem: the semantic channel encourages coherent sentence grouping, while the entity channel favors compact spans whose entity mentions form focused evidence units.

For a candidate segment $u=(s_a,\dots,s_b)$, the semantic channel uses
the embedding resultant length
$R_u=\bigl\|\sum_{i=a}^{b}\mathbf{x}_i\bigr\|$.
Let $c_u(v)$ be the count of normalized entity name $v$ in segment $u$, $N_u=\sum_v c_u(v)$, $U_u=\{v:c_u(v)>0\}$, and $\pi_u(v)=c_u(v)/N_u$. The entity channel penalizes spans that mix many weakly related entity mentions:
\begin{equation}
    \Phi(u)
    =
    N_u \mathcal{H}(\pi_u)
    +
    \frac{|U_u|-1}{2}\log N_u ,
\label{eq:entity_penalty}
\end{equation}
where the entity penalty is set to zero when $N_u=0$. The segment reward is
\begin{equation}
    r(u)
    =
    \kappa R_u
    -
    \Phi(u)
    -
    \frac{d_{\mathrm{eff}}-1}{2}\log n .
\label{eq:segment_reward}
\end{equation}
The objective is
\begin{equation}    
\mathcal{Q}_{\mathrm{MDL}}(\mathcal{U}_p)    
=    
\sum_{u\in \mathcal{U}_p} r(u),
\label{eq:quality_objective}
\end{equation}
subject to $w_{\min}\leq \mathrm{words}(u)\leq w_{\max}$. Here $\kappa$ controls the semantic reward scale and $d_{\mathrm{eff}}$ is the effective semantic dimension used in the MDL complexity term. The derivation of this entity-aware dual-channel MDL objective is in Appendix~\ref{app:derivation_dual_mdl}.

This objective keeps semantically coherent sentences together while discouraging segments whose entity mentions are diffuse across unrelated topics. The semantic term rewards groups of sentences with consistent embedding directions, while the entity term can be viewed as the coding cost of the empirical entity-mention distribution within a segment. The MDL complexity term penalizes excessive segmentation. The resulting SUs are compact text spans suitable for retrieval that induce useful hyperedges over co-occurring entities.

\noindent\textbf{Dynamic Programming} The optimal partition is found by one-dimensional dynamic programming. Let $\mathrm{dp}[i]$ be the best score for prefix $(s_1,\dots,s_i)$:
\[
    \mathrm{dp}[i]
    =
    \max_{j\in\mathcal{J}(i)}
    \left\{
        \mathrm{dp}[j] + r(s_{j+1},\dots,s_i)
    \right\},
\]
where $\mathcal{J}(i)$ contains feasible boundaries satisfying the word-count constraints. Sentence embedding prefix sums and sentence-level entity counts allow each candidate segment to be scored from its boundary statistics, and the optimal partition is recovered by backtracking through parent pointers.

\phantomsection
\label{sec:entity_assembly}
\noindent\textbf{Hypergraph Assembly.}
After SU induction, HyperSU canonicalizes entity mentions into entity nodes using exact matching over normalized names and entity labels, optionally followed by a high-threshold embedding merge for near-identical mentions. For each semantic unit $u$, let $V_u$ be the set of entities it contains. HyperSU constructs the hypergraph $G_H=(V,E_H,H)$ by treating each SU as a hyperedge $e_u \in E_H$ over $V_u$, with incidence matrix $H[v,u]=1$ iff entity $v$ appears in $u$.

\subsection{Online Retrieval: Clue-Guided Bidirectional Expansion}
\label{sec:retrieval}
At retrieval time, HyperSU extracts query entities with the same GLiNER-based extractor and maps them to indexed canonical entities by exact name matching or embedding similarity. It then retrieves cross-passage evidence through clue-conditioned SU activation, bidirectional frontier expansion, and SU-based passage scoring. This design enables forward expansion to reach distant bridge entities while using clue gating and backward anchoring to suppress noisy propagation.

\phantomsection
\label{sec:clue_agent}
\noindent\textbf{Clue Agent and SU Activation.}
A Clue Agent decomposes the query into parallel clues $\mathcal{C}_q=\{c_1,\dots,c_M\}$. If no planner is used, the original query embedding acts as the single clue. For each SU $u$, HyperSU computes
\[
    \alpha(u)
    =
    \mathrm{clamp}\!\left(
        \frac{\max_i \mathrm{sim}(c_i,u)-\delta}{1-\delta},
        0, 1
    \right)^{\gamma},
\]
where $\mathrm{sim}(c_i,u)$ is cosine similarity, $\delta$ is the SU activation threshold, and $\gamma$ controls sharpening. The max over clues activates an SU when any clue matches it well. This provides a fault-tolerant activation mechanism: retrieval does not require all generated clues, or a complete multi-hop plan, to be correct. Even if some clues are noisy, a single well-matched clue can still activate relevant SUs and trigger expansion, which is especially useful for complex multi-hop questions that require several pieces of evidence.

\phantomsection
\label{sec:bidirectional}
\noindent\textbf{Bidirectional Frontier Expansion.}
Both directions use the same propagation rule. At hop $t$, an unactivated entity $v\notin A_{t-1}$ receives
\[
    c_t(v)
    =
    \max_{u:\,v\in V_u}
    \left[
        \alpha(u)\cdot
        \max_{z\in F_t\cap V_u} a(z)
    \right],
\]
where $F_t$ is the current frontier, $A_{t-1}$ is the activated entity set, and $a(z)$ is the entity activation score. HyperSU keeps the top-$K$ new entities and applies hop decay, $a(v)=c_t(v)\beta^t$.

For each direction, HyperSU records the best SU bridge score across hops:
\[
    b(u)
    =
    \max_t
    \left[
        \alpha(u)\cdot
        \max_{z\in F_t\cap V_u} a(z)
    \right].
\]
We compute $b^{\mathrm{fwd}}$ and $b^{\mathrm{bwd}}$ using the forward and backward frontiers, respectively.

Forward expansion starts from query-linked entities and explores multiple hops to improve recall. Backward anchoring starts from entities in top dense-retrieved passages and verifies forward-reached evidence against those candidates. SUs activated by both directions form
$\mathcal{U}_{\mathrm{meet}}
=
\mathrm{supp}(b^{\mathrm{fwd}})
\cap
\mathrm{supp}(b^{\mathrm{bwd}})$
and receive a convergence bonus:
\[
    s(u)
    =
    \begin{cases}
        b^{\mathrm{fwd}}(u)\cdot \mu, & u\in \mathcal{U}_{\mathrm{meet}}, \\[2pt]
        b^{\mathrm{fwd}}(u), & u\in \mathrm{supp}(b^{\mathrm{fwd}})\setminus \mathcal{U}_{\mathrm{meet}}.
    \end{cases}
\]
Backward-only SUs are not directly ranked; they are used only to verify and boost forward-reached evidence. This prevents dense-retrieved seeds from dominating the final ranking while still using them for candidate-evidence anchoring.

\phantomsection
\label{sec:projection}
\noindent\textbf{Evidence Projection and Passage Ranking.}
Activated SU evidence is projected back to passages. For passage $p$, let
$S_p=\{u:u\subseteq p,\;s(u)>0\}$ be its activated SUs. HyperSU scores
$p$ by the capped mean of its top-$L$ SU scores:
\[
    g(p)
    =
    \frac{1}{\min(L,|S_p|)}
    \sum_{i=1}^{\min(L,|S_p|)}
    s^{(i)}(p),
\]
where $s^{(i)}(p)$ is the i-th highest SU score in $p$. If $S_p$ is empty, we set $g(p)=0$. The top-ranked passages are passed to the reader LLM for answer generation. Dense retrieval is used only to initialize backward anchoring.

\section{Experiments}
We evaluate HyperSU on GraphRAG-Bench and standard multi-hop QA benchmarks, together with component ablations and efficiency analyses.
These experiments test whether source-grounded SU hyperedges and clue-guided bidirectional expansion improve generation accuracy and retrieval quality while keeping indexing and retrieval practical.

\subsection{Experimental Setup}

\noindent\textbf{Datasets}
We evaluate on two groups of benchmarks; their statistics are shown in Table~\ref{tab:dataset_stats}.

\noindent\textit{GraphRAG-Bench.}
GraphRAG-Bench~\citep{graphragbench2026} evaluates end-to-end RAG across two domains, \textit{Novel} and \textit{Medical}, and four task types: Fact Retrieval, Complex Reasoning, Contextual Summarization, and Creative Generation.

\noindent\textit{Multi-hop QA.}
Following~\citet{linearrag2026}, we use 1{,}000 questions each from HotpotQA~\citep{yang2018hotpotqa}, 2WikiMultiHopQA~\citep{ho2020constructing}, and MuSiQue~\citep{trivedi2022musique}, which together cover reasoning chains ranging from two to four hops.

\begin{table}[!t]
\centering
\footnotesize
\setlength{\tabcolsep}{2.2pt}
\renewcommand{\arraystretch}{0.92}
\caption{Statistics of evaluation datasets. GraphRAG-Bench spans two domains and four task types; multi-hop QA datasets contain 1{,}000 questions each. Corpus size is measured in characters.}
\label{tab:dataset_stats}
\begin{tabular*}{\columnwidth}{@{\extracolsep{\fill}}llrr@{}}
\toprule
\textbf{Subset} & \textbf{Task Type} & \textbf{\#Q} & \textbf{Corpus} \\
\midrule
\multicolumn{4}{@{}l}{\textit{GraphRAG-Bench}~\citep{graphragbench2026}} \\
\multirow{4}{*}{Novel} 
& Fact Retrieval      & 971  & \multirow{4}{*}{4.82\,M} \\ 
& Complex Reasoning   & 610  & \\ 
& Contextual Summ.    & 362  & \\ 
& Creative Generation & 67   & \\
\cmidrule(lr){1-4}
\multirow{4}{*}{Medical} 
& Fact Retrieval      & 1098 & \multirow{4}{*}{1.05\,M} \\ 
& Complex Reasoning   & 509  & \\ 
& Contextual Summ.    & 289  & \\ 
& Creative Generation & 166  & \\
\midrule
\multicolumn{4}{@{}l}{\textit{Multi-hop QA}~\citep{linearrag2026}} \\
\multirow{2}{*}{HotpotQA} 
& Bridge              & 811  & \multirow{2}{*}{6.10\,M} \\ 
& Comparison          & 189  & \\
\cmidrule(lr){1-4}
\multirow{4}{*}{2WikiMH} 
& Compositional       & 413  & \multirow{4}{*}{2.92\,M} \\ 
& Comparison          & 244  & \\ 
& Bridge Comp.        & 235  & \\ 
& Inference           & 108  & \\
\cmidrule(lr){1-4}
\multirow{3}{*}{MuSiQue} 
& 2-hop               & 518  & \multirow{3}{*}{6.36\,M} \\ 
& 3-hop               & 316  & \\ 
& 4-hop               & 166  & \\
\bottomrule
\end{tabular*}
\end{table}

\begin{table*}[t]
\centering
\begingroup
\setlength{\tabcolsep}{2.2pt}
\renewcommand{\arraystretch}{0.92}
\resizebox{\textwidth}{!}{%
\begin{tabular}{@{}l*{18}{c}@{}}
\toprule
& \multicolumn{9}{c}{\textbf{Novel Dataset}}
& \multicolumn{9}{c}{\textbf{Medical Dataset}} \\
\cmidrule(lr){2-10} \cmidrule(lr){11-19}
\textbf{Model}
& \multicolumn{2}{c}{\textbf{Fact Retr.}}
& \multicolumn{2}{c}{\textbf{Cmplx Reas.}}
& \multicolumn{2}{c}{\textbf{Ctx. Summ.}}
& \multicolumn{3}{c}{\textbf{Creative Gen.}}
& \multicolumn{2}{c}{\textbf{Fact Retr.}}
& \multicolumn{2}{c}{\textbf{Cmplx Reas.}}
& \multicolumn{2}{c}{\textbf{Ctx. Summ.}}
& \multicolumn{3}{c}{\textbf{Creative Gen.}} \\
\cmidrule(lr){2-3} \cmidrule(lr){4-5} \cmidrule(lr){6-7} \cmidrule(lr){8-10}
\cmidrule(lr){11-12} \cmidrule(lr){13-14} \cmidrule(lr){15-16} \cmidrule(lr){17-19}
& ACC & R-L & ACC & R-L & ACC & Cov & ACC & FS & Cov
& ACC & R-L & ACC & R-L & ACC & Cov & ACC & FS & Cov \\
\midrule
RAG
& 58.95 & 37.67 & 41.69 & 15.22 & 50.05 & \underline{82.84} & 41.37 & 47.19 & 37.86
& 63.26 & 29.22 & 58.16 & 13.95 & 63.50 & \underline{78.07} & 58.48 & 35.98 & 57.50 \\
\midrule
\multicolumn{19}{@{}l}{\textit{Graph-based RAG}} \\
MS-GraphRAG
& 49.25 & 26.22 & 51.09 & 24.24 & 64.35 & 75.00 & 38.80 & 55.43 & 35.49
& 38.72 & 26.89 & 46.68 & 22.10 & 42.23 & 22.97 & 53.32 & 32.38 & 39.70 \\
HippoRAG
& 52.74 & 26.53 & 38.50 & 11.10 & 48.38 & \textbf{85.55} & 38.99 & \underline{71.31} & 38.90
& 56.32 & 21.12 & 55.46 & 13.53 & 60.36 & 62.78 & 63.85 & \underline{69.53} & 65.89 \\
HippoRAG 2
& 60.25 & 31.34 & 53.16 & \textbf{33.73} & 64.35 & 70.67 & 48.39 & 49.51 & 31.23
& 66.12 & 36.46 & 61.72 & \underline{36.87} & 62.96 & 45.87 & 67.65 & 59.09 & 51.42 \\
LightRAG
& 58.09 & 35.27 & 48.85 & 24.08 & 48.87 & 63.47 & 23.66 & 57.35 & 25.02
& 63.57 & \underline{37.44} & 61.49 & 25.06 & 62.87 & 50.91 & 67.67 & \textbf{78.10} & 52.16 \\
RAPTOR
& 49.18 & 23.82 & 38.81 & 11.67 & 47.13 & 82.13 & 37.67 & 70.79 & 35.75
& 54.42 & 17.86 & 53.72 & 11.84 & 59.02 & \textbf{78.47} & 61.81 & 60.22 & 53.56 \\
GFM-RAG
& 59.23 & 32.81 & 45.91 & 25.57 & 61.18 & 71.17 & 49.13 & \textbf{71.45} & 43.98
& 63.46 & 33.83 & \underline{65.78} & 25.04 & 67.52 & 64.92 & 65.20 & 51.09 & 57.91 \\
LinearRAG
& 62.17 & 33.32 & 56.68 & 32.91 & 65.73 & 66.09 & 49.12 & 52.18 & 47.15
& \underline{67.21} & 35.94 & 64.29 & 35.23 & 67.36 & 47.82 & 67.05 & 63.80 & \underline{65.97} \\
\midrule
\multicolumn{19}{@{}l}{\textit{Hypergraph-based RAG}} \\
\rowcolor{blue!4}
HyperGraphRAG
& 63.13 & 27.38 & \underline{57.41} & 31.02 & 69.13 & 61.88 & \underline{53.29} & 44.97 & \underline{52.79}
& 62.73 & 21.98 & 65.40 & 31.57 & 67.92 & 75.43 & 62.14 & 48.12 & 46.41 \\
\rowcolor{blue!4}
Hyper-RAG
& \underline{65.29} & \underline{37.72} & 56.52 & 29.18 & \underline{71.83} & 69.41 & 49.92 & 54.84 & \textbf{55.71}
& 61.42 & 25.25 & 60.12 & 25.98 & \underline{72.53} & 71.85 & \underline{67.92} & 56.88 & 50.05 \\
\rowcolor{gray!10}
HyperSU (Ours)
& \textbf{67.42} & \textbf{42.19} & \textbf{61.79} & \underline{33.36} & \textbf{75.04} & 82.49 & \textbf{56.94} & 63.93 & 48.29
& \textbf{68.49} & \textbf{42.06} & \textbf{75.03} & \textbf{38.74} & \textbf{75.01} & 76.72 & \textbf{69.92} & 64.19 & \textbf{67.21} \\
\bottomrule
\end{tabular}%
}
\endgroup
\vspace{-0.1cm}
\caption{Generation quality on GraphRAG-Bench. All results are obtained from our local reproduction under the unified evaluation protocol.}
\label{tab:gen-quality}
\vspace{-0.35cm}
\end{table*}

\begin{table*}[t]
\centering
\begingroup
\setlength{\tabcolsep}{2.6pt}
\renewcommand{\arraystretch}{0.92}
\resizebox{\textwidth}{!}{%
\begin{tabular}{@{}l*{16}{c}@{}}
\toprule
& \multicolumn{8}{c}{\textbf{Novel Dataset}}
& \multicolumn{8}{c}{\textbf{Medical Dataset}} \\
\cmidrule(lr){2-9} \cmidrule(lr){10-17}
\textbf{Model}
& \multicolumn{2}{c}{\textbf{Fact Retr.}}
& \multicolumn{2}{c}{\textbf{Cmplx Reas.}}
& \multicolumn{2}{c}{\textbf{Ctx. Summ.}}
& \multicolumn{2}{c}{\textbf{Creative Gen.}}
& \multicolumn{2}{c}{\textbf{Fact Retr.}}
& \multicolumn{2}{c}{\textbf{Cmplx Reas.}}
& \multicolumn{2}{c}{\textbf{Ctx. Summ.}}
& \multicolumn{2}{c}{\textbf{Creative Gen.}} \\
\cmidrule(lr){2-3} \cmidrule(lr){4-5} \cmidrule(lr){6-7} \cmidrule(lr){8-9}
\cmidrule(lr){10-11} \cmidrule(lr){12-13} \cmidrule(lr){14-15} \cmidrule(lr){16-17}
& Rec & Rel & Rec & Rel & Rec & Rel & Rec & Rel
& Rec & Rel & Rec & Rel & Rec & Rel & Rec & Rel \\
\midrule
RAG
& 61.38 & 74.78 & 59.37 & \underline{80.33} & 69.59 & 79.56 & 32.56 & \textbf{83.02}
& 85.48 & 63.89 & 84.88 & \textbf{83.71} & 84.00 & \underline{89.75} & 44.44 & 58.77 \\
\midrule
\multicolumn{17}{@{}l}{\textit{Graph-based RAG}} \\
MS-GraphRAG
& 61.11 & 27.29 & 72.71 & 38.70 & 81.91 & 43.35 & 53.40 & 34.75
& 38.25 & 5.67 & 61.88 & 4.22 & 60.17 & 5.22 & 67.18 & 2.75 \\
HippoRAG
& \underline{80.78} & 56.30 & \underline{87.33} & 59.03 & \underline{91.40} & 59.60 & 65.47 & 46.87
& \underline{86.59} & 52.62 & 84.25 & 41.95 & 83.25 & 49.17 & 82.44 & 44.77 \\
HippoRAG 2
& 70.94 & \textbf{80.01} & 70.44 & \textbf{86.02} & 83.31 & \textbf{88.31} & 41.96 & \underline{78.87}
& 79.30 & \textbf{87.88} & 76.70 & \underline{81.44} & 77.32 & 86.43 & 61.12 & \textbf{79.18} \\
LightRAG
& 74.35 & 33.34 & 85.77 & 37.48 & 86.93 & 37.81 & 71.55 & 38.09
& 80.40 & 41.09 & 82.44 & 43.13 & 85.23 & 42.87 & 80.62 & 45.28 \\
RAPTOR
& 62.02 & 54.30 & 68.39 & 60.88 & 75.21 & 63.22 & 58.28 & 58.76
& 86.15 & 68.84 & \underline{89.44} & 53.50 & \underline{89.73} & 59.26 & 72.91 & 52.49 \\
GFM-RAG
& 75.37 & 47.93 & 78.77 & 47.05 & 87.82 & 52.91 & \underline{72.30} & 47.59
& 86.37 & 57.37 & 83.60 & 35.42 & 76.87 & 40.71 & 81.92 & 22.15 \\
LinearRAG
& 73.75 & 72.87 & 81.03 & 79.82 & 84.71 & 61.03 & 67.15 & 71.27
& 85.81 & \underline{81.74} & 84.13 & 78.17 & 87.92 & 88.09 & \underline{87.01} & 71.02 \\
\midrule
\multicolumn{17}{@{}l}{\textit{Hypergraph-based RAG}} \\
\rowcolor{blue!4}
HyperGraphRAG
& 76.57 & 71.35 & 82.45 & 60.03 & 81.92 & 62.18 & 64.32 & 68.13
& 72.13 & 58.33 & 87.52 & 72.82 & 80.09 & 72.30 & 69.91 & 72.19 \\
\rowcolor{blue!4}
Hyper-RAG
& 80.06 & 73.59 & 80.96 & 61.68 & 85.77 & 63.27 & 63.37 & 69.60
& 70.80 & 57.91 & 80.03 & 72.11 & 82.82 & 74.18 & 68.94 & 73.28 \\
\rowcolor{gray!10}
HyperSU (Ours)
& \textbf{83.17} & \underline{78.45} & \textbf{89.30} & 67.38 & \textbf{94.04} & \underline{88.19} & \textbf{78.52} & 76.34
& \textbf{88.90} & 77.52 & \textbf{92.58} & 70.14 & \textbf{93.73} & \textbf{90.04} & \textbf{93.56} & \underline{74.93} \\
\bottomrule
\end{tabular}%
}
\endgroup
\caption{Retrieval quality on GraphRAG-Bench. All results are obtained from our local reproduction under the unified evaluation protocol.}
\label{tab:ret-quality}
\vspace{-0.45cm}
\end{table*}

\noindent\textbf{Baselines}
We compare HyperSU with three groups of methods.
\textit{Standard RAG}: dense retrieval of Top-$K$ passages.
\textit{Graph-based RAG}: MS-GraphRAG~\citep{edge2024graphrag}, HippoRAG~\citep{gutierrez2024hipporag}, HippoRAG~2~\citep{gutierrez2025hipporag2}, LightRAG~\citep{guo-etal-2025-lightrag}, RAPTOR~\citep{sarthi2024raptor}, LinearRAG~\citep{linearrag2026}, KGP~\citep{wang2024kgp}, G-retriever~\citep{he2024gretriever}, E$^2$GraphRAG~\citep{zhao2025e2graphrag}, and GFM-RAG~\citep{luo2025gfmrag}.
\textit{Hypergraph-based RAG}: HyperGraphRAG~\citep{luo2025hypergraphrag} and Hyper-RAG~\citep{feng2026hyperrag}.

\noindent\textbf{Evaluation Metrics}
For GraphRAG-Bench, we follow the official task-specific protocol:
Accuracy and ROUGE-L for Fact Retrieval and Complex Reasoning,
Accuracy and Coverage Score for Contextual Summarization,
Accuracy, Faithfulness Score, and Coverage Score for Creative Generation,
and Context Recall and Context Relevance for retrieval.
For multi-hop QA, we report Contain Accuracy and GPT-evaluated Accuracy
following~\citet{wang2025archrag}. To mitigate potential dependence on
LLM-as-judge evaluation, Appendix~\ref{app:non_llm_multihop} further
reports deterministic metrics, including answer token-level F1 and
evidence Recall@5. Main-table results of HyperSU, HyperGraphRAG, and Hyper-RAG are means over 9 runs; Appendix~\ref{app:significance_test} reports their standard deviations and significance tests.

\noindent\textbf{Implementation Details} 
We rerun all baselines in our environment using the released implementations when available, under the same benchmark protocol as HyperSU. We keep the official dataset splits, corpora, task-specific metrics, reader LLM, reader prompt, the number of retrieved passages given to the reader, judge model, and evaluation scripts fixed across methods. Each method retains its official indexing, retrieval, and ranking procedure before the final reader-context selection, so the comparison standardizes answer generation and evaluation while preserving method-specific retrieval designs. Unless otherwise specified, all reader generations use temperature 0. Appendix~\ref{app:implementation} provides the full implementation details and hyperparameter settings. To validate our reproduction, we compare our results with the scores originally reported in the corresponding baseline papers or benchmark paper. The reproduced scores generally match those reported in the original papers, and the observed deviations do not change the main rankings or conclusions.

\subsection{Performance on GraphRAG-Bench}

Tables~\ref{tab:gen-quality} and~\ref{tab:ret-quality} report generation and retrieval results on GraphRAG-Bench.

\noindent\textbf{Answer accuracy} HyperSU obtains the highest ACC among the compared methods on all eight domain--task settings.
The gains are modest on Fact Retrieval, but become larger on reasoning-intensive tasks such as Complex Reasoning, Contextual Summarization, and Creative Generation. 
These results are consistent with our motivation: source-grounded SU hyperedges preserve local multi-entity evidence, while bidirectional expansion helps compose such evidence across passages.
Hypergraph-based baselines are often stronger than standard graph-based methods, suggesting the usefulness of higher-order structure; HyperSU further improves over them by avoiding LLM-generated relational summaries and using clue-guided retrieval.

\noindent\textbf{Retrieval quality} HyperSU obtains the highest Context Recall on all eight domain--task settings. Context Relevance is more mixed: conservative methods such as dense retrieval or HippoRAG~2 can achieve higher relevance on some tasks. Nevertheless, HyperSU maintains competitive relevance while substantially improving recall, suggesting that forward expansion helps reach distant evidence and backward anchoring with convergence verification helps control noise.

\subsection{Results on Multi-hop QA Benchmarks}
We further include several RAG frameworks specifically designed for multi-hop question answering and evaluate them on multi-hop QA datasets. Table~\ref{tab:multi_hop} shows that HyperSU achieves the best GPT-Acc.\ on all three datasets, improving over the strongest baseline by +1.3 on HotpotQA, +3.2 on 2WikiMultiHopQA, and +2.7 on MuSiQue. The gains are more pronounced on 2WikiMultiHopQA and MuSiQue, where longer reasoning chains require stronger evidence composition. HyperSU is slightly below LinearRAG on 2WikiMultiHopQA under Contain-Acc.; however, HyperSU still achieves higher GPT-Acc., suggesting that exact string matching may underestimate performance on semantically correct answers with different surface forms.

\begin{table}[t]
\centering
\footnotesize
\setlength{\tabcolsep}{2.2pt}
\renewcommand{\arraystretch}{0.92}
\begin{tabular*}{\columnwidth}{@{\extracolsep{\fill}}lcccccc@{}}
\toprule
\multirow{2}{*}{\textbf{Method}} 
& \multicolumn{2}{c}{\textbf{HotpotQA}} 
& \multicolumn{2}{c}{\textbf{2Wiki}} 
& \multicolumn{2}{c}{\textbf{MuSiQue}} \\
\cmidrule(lr){2-3} \cmidrule(lr){4-5} \cmidrule(lr){6-7}
& C. & G. & C. & G. & C. & G. \\
\midrule
\multicolumn{7}{@{}l}{\textit{Graph-based RAG}} \\
KGP                     & 61.7 & 61.0 & 31.2 & 30.1 & 25.4 & 29.8 \\
MS-GraphRAG             & 53.1 & 53.4 & 48.1 & 29.6 & 21.1 & 19.6 \\
G-retriever             & 42.9 & 41.2 & 46.9 & 27.5 & 13.6 & 15.9 \\
RAPTOR                  & 54.9 & 57.1 & 50.9 & 42.2 & 24.1 & 28.8 \\
$\mathrm{E}^2$GraphRAG
                         & 60.5 & 63.7 & 54.4 & 38.3 & 23.1 & 25.2 \\
LightRAG                & 59.8 & 59.3 & 55.8 & 40.2 & 27.2 & 28.2 \\
HippoRAG                & 57.2 & 59.4 & 66.3 & 60.4 & 29.7 & 24.0 \\
GFM-RAG                 & 63.5 & 66.5 & 66.3 & 58.8 & 29.6 & 34.0 \\
HippoRAG 2              & 62.3 & 64.7 & 62.9 & 55.1 & 31.1 & 35.4 \\
LinearRAG               & 64.4 & \underline{66.8} & \textbf{70.2} & 63.5 & \underline{33.7} & 36.7 \\
\midrule
\multicolumn{7}{@{}l}{\textit{Hypergraph-based RAG}} \\
\rowcolor{blue!4}
HyperGraphRAG           & 63.9 & 64.9 & 68.8 & 64.1 & 32.7 & \underline{37.5} \\
\rowcolor{blue!4}
Hyper-RAG               & \underline{65.2} & 65.6 & 66.3 & \underline{69.1} & 32.1 & 36.4 \\
\midrule
\rowcolor{gray!10}
\textbf{HyperSU} 
                         & \textbf{66.0} & \textbf{68.1} 
                         & \underline{69.2} & \textbf{72.3} 
                         & \textbf{35.0} & \textbf{40.2} \\
\bottomrule
\end{tabular*}
\vspace{-0.25em}
\caption{Multi-hop QA results. C./G. denote Contain-/GPT-Acc. All results are obtained from our local reproduction under the unified evaluation protocol. Best and second-best results are bolded and underlined.}
\label{tab:multi_hop}
\vspace{-0.35em}
\end{table}

\subsection{Component-wise Contribution}

\begin{figure}[t]
    \centering
    \vspace{-0.35em}
    \includegraphics[width=0.96\linewidth,trim=2 2 2 2,clip]{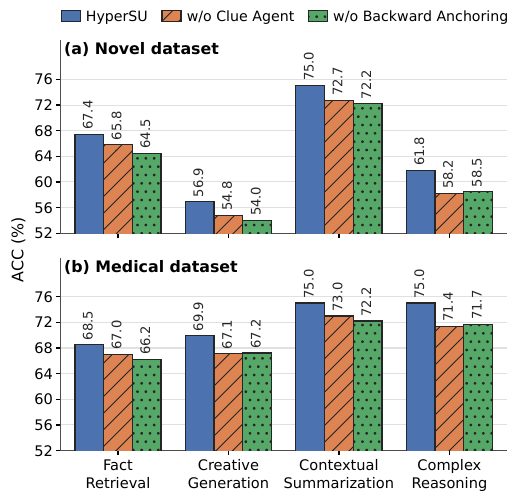}
    \vspace{-0.5em}
    \caption{HyperSU component ablation on GraphRAG-Bench; largest drops occur on Complex Reasoning.}
    \label{fig:ablation}
    \vspace{-0.55em}
\end{figure}

We ablate two GraphRAG-Bench variants: \textbf{w/o Clue Agent}, replacing multi-clue activation with a single query vector, and \textbf{w/o Backward Anchoring}, removing backward expansion and convergence verification (Figure~\ref{fig:ablation}).

Removing the Clue Agent causes the largest drop on Complex Reasoning, where queries often require several pieces of evidence. 
This is consistent with the role of parallel clues as a noise-tolerant activation mechanism: a well-matched clue can still activate relevant SUs when other clues are noisy.
Removing backward anchoring also degrades performance across tasks, suggesting that forward-only expansion may retrieve broader evidence but can accumulate noisy SUs without candidate-evidence anchoring. Appendix~\ref{app:mechanism_ablation} isolates the propagation rule itself by replacing it with PPR on the same SU hypergraph, and reports SU-level precision for the same ablation.

\subsection{Computational Efficiency}
\label{sec:efficiency}

\begin{table}[t]
\centering
\footnotesize
\setlength{\tabcolsep}{2pt}
\renewcommand{\arraystretch}{0.92}
\begin{tabular*}{\columnwidth}{@{\extracolsep{\fill}}lrrr@{}}
\toprule
\textbf{Method} &
\shortstack{\textbf{Index LLM}\\\textbf{Tokens}} &
\shortstack{\textbf{API}\\\textbf{Cost}} &
\shortstack{\textbf{Index}\\\textbf{Time}} \\
\midrule
HyperSU (Ours)   & 0        & \$0.00 & 212 s \\
HippoRAG\,2      & 1{,}321K & \$0.47 & 850 s \\
Hyper-RAG        & 2{,}047K & \$0.54 & 1{,}085 s \\
LightRAG         & 2{,}835K & \$0.68 & 2{,}400 s \\
HyperGraphRAG    & 5{,}838K & \$1.25 & 4{,}909 s \\
\bottomrule
\end{tabular*}
\caption{Offline indexing cost on Medical corpus}
\label{tab:index_cost}
\vspace{-0.2cm}
\end{table}

\begin{table}[t]
\centering
\footnotesize
\setlength{\tabcolsep}{3pt}
\renewcommand{\arraystretch}{0.92}
\begin{tabular*}{\columnwidth}{@{\extracolsep{\fill}}lrrr@{}}
\toprule
\textbf{Method} & \textbf{Tokens} & \textbf{Lat.\ (s)} & \textbf{Med.\ Avg} \\
\midrule
HyperSU (Ours)   & 406     & 3.0  & \textbf{72.11} \\
Hyper-RAG        & 521     & 4.8  & 65.50 \\
LightRAG         & 617     & 5.0  & 63.90 \\
HippoRAG\,2      & 2{,}739 & 9.5  & 64.61 \\
HyperGraphRAG    & 3{,}757 & 22.1 & 64.55 \\
\bottomrule
\end{tabular*}
\caption{Per-query retrieval cost on Medical; tokens exclude final generation, and Med. Avg is mean accuracy. For HyperSU, the 406 tokens are the complete Clue Agent usage, 256 prompt and 150 completion tokens on average.}
\label{tab:online_cost}
\vspace{-0.4cm}
\end{table}

We profile cost on the Medical corpus ($\sim$1.05\,M characters). Tables~\ref{tab:index_cost} and~\ref{tab:online_cost} report offline indexing and per-query retrieval. A formal time and space complexity analysis is provided in Appendix~\ref{app:complexity}.

\noindent\textbf{Offline indexing} HyperSU uses no generative-LLM calls for entity extraction or hyperedge construction, relying on GLiNER-based entity extraction and MDL-based SU induction. The cost of running GLiNER is included in the indexing time reported in Table~\ref{tab:index_cost}, so what HyperSU removes is the offline LLM token and API cost, not all indexing computation. As shown in Table~\ref{tab:index_cost}, HyperSU removes offline LLM indexing tokens and API cost while reducing indexing time. This reduces the cost of indexing new corpora and adapting HyperSU to new domains.

\noindent\textbf{Online retrieval} At query time, HyperSU operates over the pre-indexed SU hypergraph. The Clue Agent first produces a compact set of query clues, after which bidirectional expansion and convergence verification are performed without additional LLM-based graph construction. The reported token cost includes all LLM usage during the retrieval stage, excluding final answer generation. The Clue Agent is optional: with it disabled, the original query embedding acts as the single clue and HyperSU performs retrieval with no online LLM call, at the accuracy cost reported in Figure~\ref{fig:ablation}. Thus, HyperSU requires no generative-LLM calls for offline indexing while maintaining practical retrieval latency and strong answer accuracy.

\section{Conclusion}

We presented HyperSU, a corpus-driven semantic-unit hypergraph framework for RAG. HyperSU induces source-grounded SU hyperedges via entity-aware MDL segmentation and retrieves evidence through clue-guided bidirectional expansion, avoiding LLM-generated hyperedges and offline LLM indexing cost. Experiments on GraphRAG-Bench and three multi-hop QA benchmarks demonstrate consistent improvements over standard, graph-based, and hypergraph-based RAG baselines, with larger gains on reasoning-intensive tasks.

\section*{Limitations}

HyperSU constructs local, source-grounded SU hyperedges rather than pre-materializing predicate-level cross-document n-ary relations. This reduces combinatorial indexing cost and improves reuse, but retrieval still requires the relevant evidence chain to be connected through the induced local SUs. Failures in entity extraction, entity canonicalization, or bridge-entity coverage may therefore prevent the system from recovering otherwise valid reasoning chains.

SU quality depends on the entity-aware dual-channel MDL segmentation. The semantic channel relies on sentence embeddings to capture local discourse coherence, while the entity channel relies on accurate sentence-level mention counts. These assumptions may be less reliable for heterogeneous corpora, highly technical domains, multilingual text, or domains with sparse or ambiguous entity mentions. More adaptive segmentation objectives and domain-aware entity extraction are natural future directions.

Although clue-guided bidirectional expansion balances recall and precision, its per-query overhead still grows with corpus size and hypergraph density, which may become a bottleneck for very large-scale or latency-sensitive deployments. HyperSU also uses several hyperparameters, and the LLM-based Clue Agent introduces an online LLM call when multi-clue planning is enabled; domain transfer may therefore require re-tuning and further robustness analysis.

Evaluation is limited to English-language QA benchmarks. Multilingual settings and other knowledge-intensive tasks remain future work.

\section*{Acknowledgments}

AI assistants were used only for language polishing of the manuscript and for
minor coding assistance. All research ideas, experimental design, results, and
analysis are the authors' own, and the authors verified all content.

\bibliography{custom}

\appendix
\section{Additional Related Work Comparison}
\label{app:method-comparison}

\begin{table*}[t]
\centering
\small
\setlength{\tabcolsep}{4pt}
\begin{tabular}{llll}
\toprule
Method & Struct. & Edge source & Retrieval \\
\midrule
MS-GraphRAG & Binary KG & LLM triples & Community detection + summaries \\
LightRAG & Binary KG & LLM triples & Keyword-guided 1-hop collection \\
RAPTOR & Tree & LLM summaries & Tree / collapsed-tree search \\
HippoRAG & Binary KG & LLM triples & PPR from query entities \\
HippoRAG 2 & Binary KG & LLM triples & PPR with passage integration \\
GFM-RAG & Binary KG & LLM triples & GNN message passing \\
LinearRAG & Relation-free graph & Entity extraction & Entity--passage bridging \\
Fast-GraphRAG & Binary KG & LLM triples & PageRank graph exploration \\
Lazy-GraphRAG & Binary KG & Deferred LLM summaries & Query-time relevance testing \\
KGP & Binary KG & LLM triples & LLM-guided graph traversal \\
G-Retriever & Binary KG & LLM triples & GNN subgraph retrieval \\
$\mathrm{E}^2$GraphRAG & Hybrid tree + graph & SpaCy entities & Entity-guided chunk lookup \\
HyperGraphRAG & Hypergraph & LLM local n-ary facts & Entity/hyperedge search \\
Hyper-RAG & Hypergraph & LLM local correlations & Hypergraph diffusion \\
HyperSU & Hypergraph & Source-grounded MDL SUs & Clue-guided bidir. expansion + SU ranking \\
\bottomrule
\end{tabular}
\caption{Comparison of representative RAG methods.}
\label{tab:comparison}
\end{table*}

Table~\ref{tab:comparison} provides a compact taxonomy of representative graph- and hypergraph-based RAG methods, comparing their structural representation, edge construction source, and retrieval mechanism.
This comparison complements the related-work discussion in the main text and highlights that HyperSU differs from prior systems by inducing source-grounded semantic-unit hyperedges and performing clue-guided bidirectional expansion, rather than relying on LLM-generated triples, summaries, or local hyperedges.

\section{Theoretical Foundations and Derivations}
\label{app:theory}

\subsection{Proof of Proposition 1}
\label{app:proof_prop1}

\textbf{Proposition 1.}
\textit{For a source-grounded multi-entity evidence unit $e$ with entity set $V_e$, representing $e$ as a hyperedge preserves the unit-level membership and provenance of its co-mentioned entities. In contrast, projecting $e$ onto pairwise entity--entity edges may lose the original group identity unless auxiliary event or provenance nodes or labels are added.}

\begin{proof}
Let $V$ be a universe of entities, and let an evidence unit be represented by a subset $h \subseteq V$ with $|h|\geq 2$. A collection of evidence units is denoted by $\mathcal{E}\subseteq \mathcal{P}(V)$.

A hypergraph encoding maps $\mathcal{E}$ to $G_H=(V_{\mathcal{E}},E_H)$, where
\[
V_{\mathcal{E}}=\bigcup_{h\in\mathcal{E}}h
\qquad\text{and}\qquad
E_H=\mathcal{E}.
\]
Thus, each evidence unit is retained as a distinct hyperedge, and its group identity is directly recoverable from $E_H$.

Now consider the standard pairwise projection of $\mathcal{E}$:
\begin{equation}
\begin{aligned}
\pi(\mathcal{E})
=
\{\, \{u,v\} \mid\;& u\neq v, \\
& \exists h\in\mathcal{E}\ \mathrm{s.t.}\ 
\{u,v\}\subseteq h \,\}.
\end{aligned}
\label{eq:app_pairwise_projection}
\end{equation}
This projection is not injective. For three distinct entities $a,b,c\in V$, define
\[
\begin{aligned}
\mathcal{E}_1&=\big\{\{a,b,c\}\big\},\\
\mathcal{E}_2&=\big\{\{a,b\},\{a,c\},\{b,c\}\big\}.
\end{aligned}
\]
The first collection contains one three-entity evidence unit, while the second contains three separate pairwise units. However,
\[
\pi(\mathcal{E}_1)=\pi(\mathcal{E}_2)
=
\big\{\{a,b\},\{a,c\},\{b,c\}\big\}.
\]
Therefore, after pairwise projection, one cannot determine whether the three entities appeared in a single shared evidence unit or in three independent pairwise units.

The same argument holds for any evidence unit $h$ with $|h|\geq 3$: the single hyperedge $\{h\}$ and the set of all pairwise edges $\binom{h}{2}$ yield the same pairwise projection but encode different evidence-unit memberships. If each evidence unit is associated with a source span or provenance identifier, this group-level provenance is also not recoverable from unlabeled pairwise edges alone. Hence, pairwise projection loses group identity and provenance. A binary graph can recover this information by introducing auxiliary event, document, or provenance nodes, but such nodes are additional modeling structure beyond the plain pairwise entity projection considered here.
\end{proof}

\subsection{Equivalence to the Entity--SU Incidence Bipartite Graph}
\label{app:star_expansion}

Proposition~1 contrasts the hyperedge representation with the pairwise
entity--entity projection. A different encoding, the entity--SU incidence
bipartite graph, preserves the same information as the hypergraph, and we
state the correspondence here.

Let $G_H=(V,E_H,H)$ be the SU hypergraph of \S\ref{sec:entity_assembly}.
Star expansion~\citep{agarwal2006higherorder,zhou2006hypergraph,sun2008hypergraph}
replaces each SU hyperedge $e_u\in E_H$, with entity set $V_u$, by an
auxiliary SU node $h_u$ connected to every $v\in V_u$. The resulting
bipartite graph $B$ has biadjacency matrix
\[
    B[v,h_u]=H[v,e_u]=1 \quad\text{iff}\quad v\in V_u .
\]
As long as the two node types are preserved, the correspondence is
bijective: the neighborhood of each SU node recovers its hyperedge, and each
hyperedge determines its SU node and incident edges. The encoding is
therefore lossless in both directions, so the entity--SU bipartite graph and
the SU hypergraph are equivalent representations of the same incidence
structure. A provenance-node graph is the same construction with $h_u$ read
as a provenance node.

\subsection{Derivation of the Entity-Aware Dual-Channel MDL Objective}
\label{app:derivation_dual_mdl}

This section derives the objective in Eq.~\ref{eq:quality_objective}. Let $p=(s_1,\dots,s_n)$ be a passage with L2-normalized sentence embeddings $X=(\mathbf{x}_1,\dots,\mathbf{x}_n)$, and let $\mathcal{U}_p=\{u_1,\dots,u_K\}$ be a contiguous non-overlapping partition. HyperSU scores each candidate SU with a semantic channel and an entity channel under the minimum description length (MDL) principle~\citep{rissanen1978mdl,grunwald2007mdl}.

\paragraph{Semantic channel.}
For a segment $u$, the semantic channel uses the embedding resultant length
\[
R_u=\Bigl\|\sum_{i\in u}\mathbf{x}_i\Bigr\|.
\]
This can be viewed as the sufficient statistic of a von Mises--Fisher-style directional model on normalized sentence embeddings~\citep{banerjee2005vmf}. Up to constants that do not depend on the partition, the semantic reward is $\kappa R_u$, where $\kappa$ controls the semantic reward scale.

\paragraph{Entity channel.}
GLiNER provides sentence-level entity mentions before SU induction. For a candidate segment $u$, let $c_u(v)$ be the count of normalized entity name $v$, $N_u=\sum_v c_u(v)$, $U_u=\{v:c_u(v)>0\}$, and $\pi_u(v)=c_u(v)/N_u$. If $N_u>0$, the negative log-likelihood of the entity mention multiset under its segment-level multinomial MLE is
\begin{equation}
N_u\mathcal{H}(\pi_u)
=
-\sum_{v\in U_u} c_u(v)\log \pi_u(v).
\label{eq:app_entity_nll}
\end{equation}
A BIC-style code length for the entity distribution contributes~\citep{schwarz1978bic}
\[
\frac{|U_u|-1}{2}\log N_u.
\]
When $N_u=0$, both entity terms are set to zero.

\paragraph{Model complexity and objective.}
The semantic channel also contributes a segment-level complexity term using an effective semantic dimension $d_{\mathrm{eff}}$:
\[
\frac{d_{\mathrm{eff}}-1}{2}\log n.
\]
Combining the semantic reward, entity data code, and complexity terms yields the segment reward
\begin{equation}
\begin{aligned}
r(u)
={}&
\kappa R_u
-
N_u\mathcal{H}(\pi_u)
-
\frac{|U_u|-1}{2}\log N_u  \\
&-
\frac{d_{\mathrm{eff}}-1}{2}\log n,
\end{aligned}
\label{eq:app_segment_reward}
\end{equation}
with the entity terms omitted when $N_u=0$. The full objective is segment-additive:
\begin{equation}
\mathcal{Q}_{\mathrm{MDL}}(\mathcal{U}_p)
=
\sum_{u\in \mathcal{U}_p} r(u),
\label{eq:app_mdl_objective}
\end{equation}
subject to the word-count constraints
\[
w_{\min}\leq \mathrm{words}(u)\leq w_{\max}.
\]

The semantic channel favors merging directionally coherent adjacent spans, whereas the entity channel can prevent merges that mix unrelated entity contexts. Thus, the objective produces SUs that are coherent as text spans and compact as entity-bearing hyperedges.

\subsection{Proof of Dynamic Programming Optimality}
\label{app:proof_prop2}

\textbf{Proposition 2.}
\textit{For a passage of $n$ sentences with pre-computed candidate-segment rewards, the optimal SU partition maximizing $\mathcal{Q}_{\mathrm{MDL}}$ under contiguity, non-overlap, and word-count constraints can be found exactly by dynamic programming.}

\begin{proof}
The objective is segment-additive. For any contiguous non-overlapping partition $\mathcal{U}_p$,
\[
\mathcal{Q}_{\mathrm{MDL}}(\mathcal{U}_p)
=
\sum_{u\in \mathcal{U}_p}r(u),
\]
where $r(u)$ depends only on the boundary statistics of segment $u$. Let $\mathcal{Q}^*(i)$ denote the optimal objective value for prefix $(s_1,\dots,s_i)$. If the last segment of an optimal partition ends at $i$ and starts at $j+1$, then the preceding prefix $(s_1,\dots,s_j)$ must itself be optimally partitioned. Otherwise, replacing it with a better prefix partition would improve the full objective.

This yields the Bellman recurrence
\begin{equation}
\mathcal{Q}^*(i)
=
\max_{j\in\mathcal{J}(i)}
\bigl\{\mathcal{Q}^*(j)+r(j,i)\bigr\},
\label{eq:app_dp_recurrence}
\end{equation}
where $\mathcal{J}(i)$ is the set of feasible previous boundaries under the word-count constraints, and $r(j,i)$ is the dual-channel segment reward for $(s_{j+1},\dots,s_i)$. Setting $\mathrm{dp}[0]=0$ and computing $\mathrm{dp}[i]=\mathcal{Q}^*(i)$ for $i=1,\dots,n$ gives the global optimum at $\mathrm{dp}[n]$. Parent pointers recover the corresponding SU partition.

Embedding prefix sums compute $R_u$ from segment boundaries, and sentence-level entity-count statistics compute the entity terms. Since the recurrence enumerates all feasible last-segment choices and combines them with optimal prefix solutions, the recovered partition is globally optimal.
\end{proof}

\subsection{Overall Complexity Analysis}
\label{app:complexity}

Let \(P\) be the number of passages, \(n_p\) the number of sentences in passage \(p\), 
\(N=\sum_{p=1}^{P} n_p\), \(U=|\mathcal{E}_H|\) the number of induced semantic units 
or hyperedges, \(V=|\mathcal{V}|\) the number of canonical entities, and 
\(I=\sum_{u\in \mathcal{E}_H}|V_u|\) the number of entity--SU incidences. Let \(d\) be 
the embedding dimension, \(M\) the number of query clues, \(T_{\mathrm{fwd}}\) and 
\(T_{\mathrm{bwd}}\) the forward and backward expansion depths, and \(K\) the number 
of retained entities per hop.

\noindent\textbf{Offline indexing.}
Excluding neural encoder and GLiNER inference, the main algorithmic cost comes from 
entity-aware MDL segmentation and hypergraph assembly. For passage \(p\), let \(C_p\) 
be the number of feasible candidate segments satisfying the word-count constraints. 
Using sentence-embedding prefix sums and sentence-level entity-count statistics, each 
candidate segment can be scored from boundary statistics, and the one-dimensional 
dynamic program considers all feasible previous boundaries. Therefore, the total 
offline segmentation cost is
\[
O\left(\sum_{p=1}^{P} C_p(d+\bar{m}_p)\right),
\]
where \(\bar{m}_p\) denotes the average number of distinct entity types touched when 
scoring a candidate segment. In the worst case, \(C_p=O(n_p^2)\). In practice, because 
the word-count constraints bound the maximum segment length and passages are bounded 
chunks, \(C_p\) is near-linear in \(n_p\). Entity canonicalization by normalized exact 
matching and sparse incidence construction are linear in the number of extracted 
mentions and incidences, giving \(O(I)\) additional time and memory.

\noindent\textbf{Online retrieval.}
For each query, clue-conditioned SU activation over all SUs costs \(O(MUd)\) with 
exhaustive similarity computation. Bidirectional frontier expansion then visits only 
the local neighborhoods of retained frontier entities:
\[
O\left(
\sum_{t=1}^{T_{\mathrm{fwd}}+T_{\mathrm{bwd}}}
\sum_{z\in F_t}\sum_{u:z\in V_u}|V_u|
\right).
\]
With top-\(K\) pruning, average incident-SU degree \(\bar{h}\), and average hyperedge 
cardinality \(\bar{r}\), this becomes 
\(O((T_{\mathrm{fwd}}+T_{\mathrm{bwd}})K\bar{h}\bar{r})\). Projecting activated SUs 
back to passages is linear in the number of activated SUs. Thus, with fixed 
\(M\), \(T_{\mathrm{fwd}}\), \(T_{\mathrm{bwd}}\), \(K\), and capped-mean parameter 
\(L\), the online graph traversal scales with the size and density of the activated 
local neighborhood, while the main corpus-level term is SU activation.

\section{Implementation and Reproducibility Details}
\label{app:implementation}

This section reports implementation details that are not expanded in the
main text. HyperSU uses GLiNER for offline entity extraction. LLM calls
are used for the Clue Agent, answer generation, and GPT-Acc evaluation.

\subsection{Experimental Protocol Details}
\label{app:experimental_protocol}

\noindent\textbf{Unified reproduction protocol.}
All baseline results reported in the main tables are reproduced in our environment. 
For each benchmark, we keep the official question splits, corpora, task-specific metrics, final reader LLM, reader prompt, final top-$k$ reader-context budget, judge model, and evaluation scripts fixed across methods. 
Each method retains its own official indexing, retrieval, and ranking procedure before the final reader-context selection, so the comparison standardizes answer generation and evaluation while preserving method-specific retrieval designs. 
We use released implementations when available and faithful reimplementations otherwise, following official configurations or paper-reported defaults. 
For all reproduced methods, we pass the top-5 retrieved passages to the final answer generator and do not add an extra reranker unless it is part of the released method itself. 
Unless otherwise specified, all LLM calls are run with temperature $0$. All local indexing, retrieval, reproduction, and efficiency-profiling runs are conducted on the machine configuration reported in Table~\ref{tab:machine_config}.

\begin{table}[t]
\centering
\small
\setlength{\tabcolsep}{3pt}
\renewcommand{\arraystretch}{0.95}
\caption{Machine configuration used in our experiments.}
\label{tab:machine_config}
\begin{tabularx}{\columnwidth}{@{}lX@{}}
\toprule
\textbf{Component} & \textbf{Specification} \\
\midrule
GPU & NVIDIA RTX PRO 6000 Blackwell Workstation Edition, 96\,GB VRAM \\
CPU & AMD Ryzen 9 9950X3D, 16 cores / 32 threads \\
Memory & 91 GiB RAM \\
OS & Ubuntu 22.04.4 LTS \\
CUDA & CUDA 13.1 driver support; CUDA Toolkit 12.4 \\
\bottomrule
\end{tabularx}
\end{table}

\begin{table*}[t]
\centering
\small
\setlength{\tabcolsep}{4pt}
\caption{GLiNER label sets used by HyperSU.}
\label{tab:gliner_label_sets}
\begin{tabular}{@{}p{0.15\textwidth}p{0.78\textwidth}@{}}
\toprule
\textbf{Label Set} & \textbf{Labels} \\
\midrule
Medical &
\texttt{disease}, \texttt{disorder}, \texttt{symptom},
\texttt{clinical sign}, \texttt{drug}, \texttt{treatment},
\texttt{procedure}, \texttt{body part}, \texttt{organ},
\texttt{diagnostic test}, \texttt{lab test} \\
\midrule
General &
\texttt{person}, \texttt{character}, \texttt{book}, \texttt{novel},
\texttt{film}, \texttt{song}, \texttt{work of art}, \texttt{location},
\texttt{country}, \texttt{city}, \texttt{organization}, \texttt{event},
\texttt{date}, \texttt{year}, \texttt{object}, \texttt{product},
\texttt{scientific concept} \\
\bottomrule
\end{tabular}
\end{table*}

\begin{table*}[t]
\centering
\small
\setlength{\tabcolsep}{6pt}
\caption{Main hyperparameter settings of HyperSU.}
\label{tab:main_hyperparameters}
\begin{tabular}{@{}lll@{}}
\toprule
\textbf{Component} & \textbf{Setting} & \textbf{Value} \\
\midrule
Entity extraction & GLiNER model & \texttt{urchade/gliner\_large-v2.1} \\
Entity extraction & confidence threshold & $0.3$ \\
Entity canonicalization & matching rule & normalized exact match \\
Entity canonicalization & embedding merge & disabled in main experiments \\
Embedding & embedding model & \texttt{BAAI/bge-large-en-v1.5} \\
SU induction & semantic reward scale & $\kappa=75$ \\
SU induction & effective semantic dimension & $d_{\mathrm{eff}}=32$ \\
Clue Agent & maximum clues & $M=3$ \\
SU activation & activation threshold & $\delta=0.5$ \\
SU activation & sharpening factor & $\gamma=1.0$ \\
Forward expansion & depth & $T^{\mathrm{fwd}}=4$ \\
Forward expansion & candidates per hop & $30$ \\
Forward expansion & hop decay & $\beta=0.5$ \\
Backward anchoring & depth & $T^{\mathrm{bwd}}=2$ \\
Backward anchoring & dense seeds & top-10 passages \\
Convergence verification & convergence bonus & $\mu=2.0$ \\
Passage projection & capped mean & $L=3$ \\
Reader context & final passages & top-5 passages \\
Post-retrieval ranking & reranker & none \\
LLM inference & temperature & $0$ \\
Main LLM & Clue Agent and reader & GPT-4o-mini \\
\bottomrule
\end{tabular}
\end{table*}

\subsection{Entity Extraction and Canonicalization Details}
\label{app:entity_extraction_details}

HyperSU uses \texttt{urchade/gliner\_large-v2.1} for sentence-level
entity extraction with confidence threshold $0.3$. Each extracted
mention keeps its surface span, entity label, confidence score,
sentence ID, and passage ID. Mention names are normalized by
lowercasing, Unicode normalization, punctuation cleanup, and low-value
mention filtering.

We use two GLiNER label sets, shown in
Table~\ref{tab:gliner_label_sets}. The medical label set is used for
the Medical domain, and the general label set is used for the Novel
domain and multi-hop QA datasets. At query time, extracted query
entities are normalized with the same rules as corpus entities and are
linked to indexed canonical entities by normalized exact matching. The
implementation supports optional high-threshold embedding-based merging
for near-identical entity names, but this option is not enabled in the
main experiments.

\subsection{Hyperparameter Settings}
\label{app:hyperparameter_settings}

Table~\ref{tab:main_hyperparameters} summarizes the main HyperSU
settings used in the main experiments. Unless otherwise specified,
clue-conditioned SU activation uses a fixed global threshold
$\delta=0.5$ and sharpening factor $\gamma=1.0$ across all datasets.
Appendix~E further analyzes the sensitivity of retrieval performance
to activation and expansion hyperparameters.

\subsection{Prompt A: Clue Agent}
\label{app:prompt_clue}

\begin{promptbox}
\textit{Used by the Clue Agent to extract independent reasoning clues
from a multi-hop query. Each clue targets a different evidence aspect
and is used in parallel during SU activation. In the main experiments,
the retriever uses at most $M=3$ valid clues.}

\tcblower

\textbf{[System]} You are a clue extraction agent that identifies the
independent evidence aspects needed to answer a multi-hop question.
Each clue should describe a distinct piece of evidence that the
retrieval system should look for.

\textbf{Hard Rules:} Return valid JSON only. Each clue MUST be a
self-contained natural-language description of one aspect of the
evidence needed. Ground each clue in entities or concepts explicitly
mentioned or directly implied by the query. Prefer the fewest clues
that cover all reasoning aspects (typically 2--4). Do NOT produce
trivial paraphrases of the original query.
\end{promptbox}

\subsection{Prompt B: GPT-Acc Evaluation}
\label{app:prompt_eval}

\begin{promptbox}
\textit{Used to compute GPT-Acc.\ on multi-hop QA benchmarks.}

\tcblower

\textbf{[System]} You are an expert evaluator.

\medskip
\textbf{[User]} Please evaluate whether the generated answer is correct
by comparing it with the gold answer.

Generated answer: \texttt{\{pre\_answer\}} \\
Gold answer: \texttt{\{gold\_ans\}}

The generated answer should be considered correct if it:
(1) contains the key information from the gold answer;
(2) is factually accurate and consistent with the gold answer; and
(3) does not contain contradicting information.

Respond with only ``correct'' or ``incorrect''.
\end{promptbox}

\noindent\textbf{Artifact Licenses and Terms of Use.}
\label{app:artifact-licenses} We use third-party artifacts only through their official public releases and follow the corresponding licenses, access conditions, and terms of use. The benchmark datasets used in this paper, including GraphRAG-Bench, HotpotQA, 2WikiMultiHopQA, and MuSiQue, are used solely for research evaluation. We do not redistribute the raw benchmark data; instead, our release provides code, configuration files, prompts, and preprocessing instructions that allow users to obtain the data from the official sources. We also use publicly released model and software artifacts, including the GLiNER entity extractor, the embedding model, baseline implementations, and evaluation scripts, under their respective licenses. We preserve copyright and license notices for third-party code and require users of our released code to comply with the original licenses of these dependencies. API-based language models used for clue generation, answer generation, or evaluation are accessed under the provider's terms of service, and no proprietary model weights are redistributed. Our own HyperSU implementation is released under MIT. Any derived indices, cached outputs, or processed artifacts are released only when permitted by the licenses or terms of the underlying datasets and models.

\subsection{Data Considerations}
\label{app:data-considerations}

We do not collect new data. All experiments use public benchmarks as
released, and we rely on the filtering their creators report. The
multi-hop QA corpora are built from Wikipedia and therefore name public
figures; we use this text only as retrieval evidence and do not attempt
to re-identify anyone. Because semantic units are verbatim source spans,
HyperSU introduces no personal information of its own, but it can surface
whatever a corpus already contains, so deployments over sensitive corpora
need the same access controls as the corpus itself. We did not run an
additional offensive-content audit.

\begin{table*}[t]
\centering
\small
\setlength{\tabcolsep}{6pt}
\caption{Controlled comparison of hyperedge construction strategies on
GraphRAG-Bench, averaged over the eight domain--task settings. All
variants use the same downstream entity canonicalization, Clue Agent,
bidirectional expansion, SU-to-passage projection, and generation LLM.
Only the offline hyperedge construction strategy is changed.}
\label{tab:hyperedge_construction_avg}
\begin{tabular}{lccc}
\toprule
\textbf{Method} & \textbf{ACC} & \textbf{Context Recall} & \textbf{Context Relevance} \\
\midrule
Sentence-level hyperedge & 63.42 & 83.07 & 68.4 \\
Fixed-window hyperedge (3 sent.) & 65.18 & 85.11 & 71.1 \\
Semantic chunking hyperedge & 64.37 & 86.24 & 73.9 \\
MDL w/o entity channel & 65.29 & 84.78 & 69.7 \\
MDL w/o semantic channel & 65.36 & 85.62 & 70.2 \\
HyperGraphRAG-style generated hyperedges & 64.92 & 85.01 & 74.5 \\
HyperRAG-style generated hyperedges & 61.18 & 82.36 & 73.6 \\
\textbf{Entity-aware MDL (ours)} & \textbf{68.71} & \textbf{89.23} & \textbf{77.9} \\
\bottomrule
\end{tabular}
\end{table*}

\begin{table*}[t]
\centering
\small
\setlength{\tabcolsep}{4pt}
\caption{LLM-based evaluation of hyperedge representation quality using
GPT-4o-mini. Scores are on a 1--5 scale. ``Overall'' is the average of
faithfulness, information preservation, self-containedness, and retrieval
usefulness. ``MDL Win Rate'' denotes the blind pairwise preference rate
of entity-aware MDL against each baseline.}
\label{tab:hyperedge_quality_llm}
\begin{tabular}{lcccccc}
\toprule
\textbf{Method} & \textbf{Faith.} & \textbf{Pres.} &
\textbf{Self-Cont.} & \textbf{Retr. Use.} &
\textbf{Overall} & \textbf{MDL Win Rate} \\
\midrule
Sentence-level hyperedge & 4.61 & 3.42 & 3.18 & 3.27 & 3.62 & 71.6\% \\
Fixed-window hyperedge (3 sent.) & 4.33 & 3.89 & 3.71 & 3.85 & 3.94 & 64.3\% \\
Semantic chunking hyperedge & 4.41 & 4.07 & 3.98 & 4.11 & 4.14 & 60.2\% \\
MDL w/o entity channel & 4.82 & 4.24 & 4.09 & 3.96 & 4.28 & 63.2\% \\
MDL w/o semantic channel & 4.77 & 4.19 & 4.25 & 4.01 & 4.31 & 59.6\% \\
HyperGraphRAG-style generated hyperedges & 4.09 & 4.31 & 4.37 & 4.19 & 4.24 & 56.9\% \\
HyperRAG-style generated hyperedges & 4.05 & 4.27 & 4.34 & 4.15 & 4.20 & 58.4\% \\
\textbf{Entity-aware MDL (ours)} & \textbf{4.91} & \textbf{4.52} &
\textbf{4.49} & \textbf{4.58} & \textbf{4.63} & -- \\
\bottomrule
\end{tabular}
\end{table*}

\section{Analysis of Semantic-Unit Hyperedge Construction}
\label{app:su_hyperedge_construction}

This section analyzes the construction of semantic-unit (SU)
hyperedges. We focus on four questions: whether the proposed
entity-aware MDL construction outperforms alternative hyperedge
construction strategies, whether the induced hyperedges are intrinsically
faithful and useful for retrieval, how the MDL segmentation
hyperparameters affect downstream performance, and what segmentation
behavior the MDL objective produces in practice.

\subsection{Controlled Comparison of Hyperedge Construction Strategies}
\label{app:hyperedge_construction}

This experiment isolates hyperedge construction from the retrieval
algorithm. \textbf{Sentence-level hyperedge} uses each sentence as one
hyperedge. \textbf{Fixed-window hyperedge} partitions a passage into
non-overlapping three-sentence windows. \textbf{Semantic chunking
hyperedge} uses adjacent embedding similarity to form variable-length
spans. \textbf{MDL w/o entity channel} removes the entity-mention coding
term and keeps only semantic cohesion and segmentation complexity.
\textbf{MDL w/o semantic channel} removes the semantic reward and keeps
only the entity channel and segmentation complexity. The two generated
baselines follow the local hyperedge generation procedures of
HyperGraphRAG and Hyper-RAG, but replace their native retrieval modules
with the HyperSU retrieval pipeline.

Table~\ref{tab:hyperedge_construction_avg} shows that entity-aware MDL
achieves the best ACC, Context Recall, and Context Relevance.
Sentence-level hyperedges are
faithful but too fragmented, while fixed-window hyperedges impose rigid
boundaries. Semantic chunking improves over these simple baselines but
does not explicitly model entity compactness. The two single-channel MDL
ablations further show that both channels are necessary. Removing the
entity channel reduces Context Recall, indicating that semantic cohesion
alone does not sufficiently control entity diffusion inside hyperedges.
Removing the semantic channel preserves relatively high recall but still
underperforms full MDL in ACC, suggesting that entity compactness alone
does not guarantee coherent, self-contained evidence spans. The full
dual-channel objective achieves the best result by jointly modeling
sentence-level coherence and entity-mention compactness.

\subsection{Intrinsic Evaluation of Hyperedge Representation Quality}
\label{app:hyperedge_quality}

We also evaluate the intrinsic quality of the hyperedge sets. For each
query--passage instance, GPT-4o-mini is shown the query, the source
passage, and the complete hyperedge set produced by one construction
strategy. The judge scores the set on faithfulness, information
preservation, self-containedness, and retrieval usefulness. We also
conduct blind pairwise comparisons between entity-aware MDL and each
baseline, with method identities hidden and output order randomized.

Table~\ref{tab:hyperedge_quality_llm} explains the performance trends in
Table~\ref{tab:hyperedge_construction_avg}. Sentence-level hyperedges
remain highly faithful because they are directly copied from source
sentences, but they are less self-contained and less useful for
multi-hop retrieval. Generated hyperedges improve preservation and
self-containedness, but their faithfulness scores are lower, reflecting
the risk of unsupported or over-abstracted generated relations. The two
single-channel MDL variants are stronger than simple segmentation
baselines, yet both are consistently below full MDL. Entity-aware MDL
achieves the best overall score by preserving source faithfulness while
forming coherent and retrieval-useful multi-entity evidence units.

\subsection{Sensitivity to MDL Segmentation Hyperparameters}
\label{app:mdl_hparam}

The entity-aware dual-channel MDL objective contains two important
segmentation hyperparameters: the semantic reward scale $\kappa$ and
the effective semantic dimension $d_{\mathrm{eff}}$. The former controls
the strength of the semantic cohesion reward, while the latter controls
the segment-level complexity penalty. Since these parameters affect SU
granularity and hyperedge structure, we analyze their impact on
GraphRAG-Bench performance.

Unless otherwise stated, all settings follow the main experiments. For
each configuration, we rebuild the SU hypergraph and run the full
retrieval pipeline. We report average ACC and Context Recall over the
eight GraphRAG-Bench domain--task settings. We also report the average
number of SUs per passage $\bar{k}$ and the average number of sentences
per SU $\bar{\ell}$ to characterize segmentation behavior.

\paragraph{Sensitivity to $\kappa$.}
Table~\ref{tab:kappa_sensitivity} shows the effect of varying $\kappa$
with $d_{\mathrm{eff}}=32$ fixed. Performance follows an inverted-U
trend. When $\kappa$ is too small, the semantic reward is weak and the
dynamic program tends to form coarse SUs, which dilute entity
co-mention signals. When $\kappa$ is too large, the semantic reward
dominates and the segmentation becomes too fine, reducing the advantage
of higher-order hyperedges. The default value $\kappa=75$ achieves the
best balance.

\begin{table}[t]
\centering
\small
\setlength{\tabcolsep}{5pt}
\caption{Sensitivity to the semantic reward scale $\kappa$ with
$d_{\mathrm{eff}}=32$. ACC and Recall are averaged over the eight
GraphRAG-Bench domain--task settings.}
\label{tab:kappa_sensitivity}
\begin{tabular}{ccccc}
\toprule
$\boldsymbol{\kappa}$ & $\boldsymbol{\bar{k}}$ &
$\boldsymbol{\bar{\ell}}$ & \textbf{ACC} & \textbf{Recall} \\
\midrule
25  & 1.42 & 11.75 & 64.28 & 81.34 \\
50  & 2.39 & 7.12  & 67.03 & 86.06 \\
\textbf{75}  & \textbf{3.12} & \textbf{5.44}  &
\textbf{68.71} & \textbf{89.23} \\
100 & 3.83 & 4.47  & 68.26 & 87.01 \\
150 & 5.20 & 3.20  & 66.17 & 84.74 \\
200 & 6.42 & 2.66  & 63.43 & 82.17 \\
\bottomrule
\end{tabular}
\end{table}

\paragraph{Sensitivity to $d_{\mathrm{eff}}$.}
Table~\ref{tab:deff_sensitivity} varies $d_{\mathrm{eff}}$ with
$\kappa=75$ fixed. The role of $d_{\mathrm{eff}}$ is not to estimate the
full embedding dimension. Instead, it acts as an effective capacity
parameter in the MDL complexity term. From the MDL perspective, which
operationalizes Occam's principle, a good segmentation should explain
local semantic coherence without paying unnecessary segmentation
complexity. When $d_{\mathrm{eff}}$ is too small, the complexity penalty
is weak and the dynamic program tends to over-segment passages. When
$d_{\mathrm{eff}}$ is too large, the penalty becomes too strong and the
model favors overly coarse segments. The stable region around
$d_{\mathrm{eff}}\in[16,64]$ suggests that HyperSU only requires a
moderate complexity-control scale.

\begin{table}[t]
\centering
\small
\setlength{\tabcolsep}{5pt}
\caption{Sensitivity to the effective semantic dimension
$d_{\mathrm{eff}}$ with $\kappa=75$. ACC and Recall are averaged over
the eight GraphRAG-Bench domain--task settings.}
\label{tab:deff_sensitivity}
\begin{tabular}{ccccc}
\toprule
$\boldsymbol{d_{\mathrm{eff}}}$ & $\boldsymbol{\bar{k}}$ &
$\boldsymbol{\bar{\ell}}$ & \textbf{ACC} & \textbf{Recall} \\
\midrule
8   & 5.67 & 3.04  & 65.73 & 84.22 \\
16  & 4.25 & 4.02  & 67.86 & 86.86 \\
\textbf{32}  & \textbf{3.12} & \textbf{5.44}  &
\textbf{68.71} & \textbf{89.23} \\
64  & 2.55 & 6.73  & 68.00 & 87.18 \\
128 & 1.91 & 8.84  & 66.48 & 85.47 \\
256 & 1.39 & 12.93 & 63.05 & 81.02 \\
\bottomrule
\end{tabular}
\end{table}

\begin{figure}[t]
    \centering
    \includegraphics[width=\linewidth]{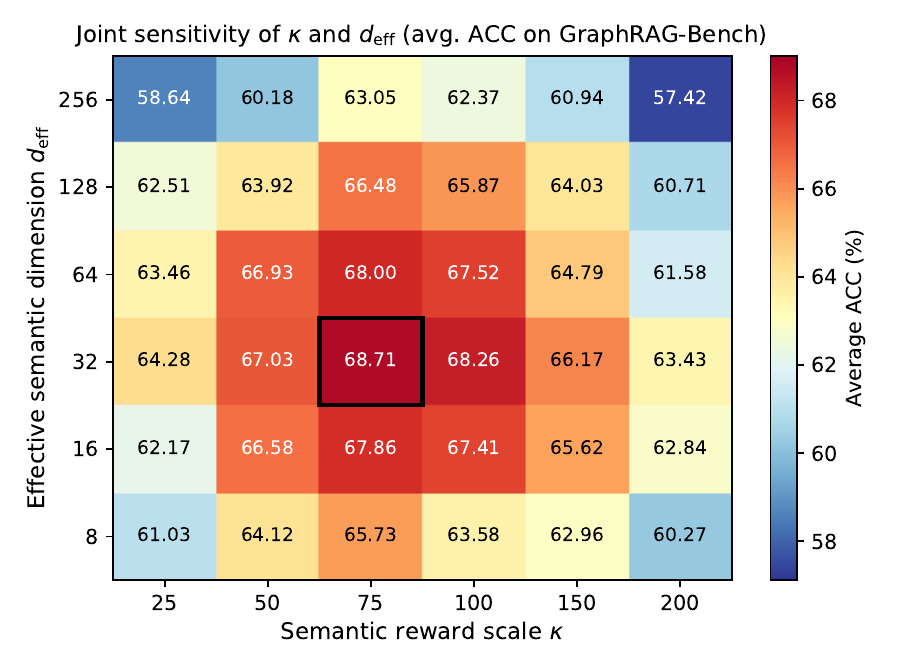}
    \caption{Joint sensitivity of $\kappa$ and $d_{\mathrm{eff}}$ on
    GraphRAG-Bench. Each cell reports average ACC over the eight
    domain--task settings. The central region is stable, while extreme
    settings lead to overly coarse or overly fine SU segmentation.}
    \label{fig:joint_mdl_sensitivity}
\end{figure}

\begin{table*}[t]
\centering
\footnotesize
\setlength{\tabcolsep}{4pt}
\renewcommand{\arraystretch}{1.12}
\caption{Examples of MDL-induced semantic units from 2WikiMultiHopQA.
Each SU is a contiguous source span and induces a hyperedge over its
canonical entity vertices.}
\label{tab:mdl_su_examples}
\begin{tabular}{@{}p{0.16\textwidth}p{0.47\textwidth}p{0.29\textwidth}@{}}
\toprule
\textbf{Type} & \textbf{Semantic Unit Text} & \textbf{Hyperedge Vertices} \\
\midrule
Musician and band affiliations &
Billy Milano is a Bronx-born heavy metal musician now based in Austin,
Texas. He is the singer and occasionally guitarist and bassist of
crossover thrash band M.O.D., and he was also the singer of its
predecessor, Stormtroopers of Death. He was also the singer of United
Forces, which also featured his Stormtroopers of Death bandmate Dan
Lilker. &
Billy Milano; Bronx; Austin; Texas; M.O.D.; Stormtroopers of Death;
United Forces; Dan Lilker \\
\midrule
Film adaptation and production &
The Private Life of Helen of Troy is a 1927 American silent film about
Helen of Troy based on the 1925 novel of the same name by John Erskine,
and adapted to screen by Gerald Duffy. The film was directed by
Alexander Korda and starred Maria Corda as Helen, Lewis Stone as
Menelaus, and Ricardo Cortez as Paris. &
The Private Life of Helen of Troy; Helen of Troy; John Erskine;
Gerald Duffy; Alexander Korda; Maria Corda; Helen; Lewis Stone;
Menelaus; Ricardo Cortez; Paris \\
\bottomrule
\end{tabular}
\end{table*}

\paragraph{Joint sensitivity.}
Figure~\ref{fig:joint_mdl_sensitivity} visualizes the joint
$\kappa\times d_{\mathrm{eff}}$ sweep on GraphRAG-Bench. The best performance
occurs around the central parameter settings rather than at the
extremes. Configurations around $\kappa\in[50,100]$ and
$d_{\mathrm{eff}}\in[16,64]$ remain close to the best score, indicating
that the MDL objective remains robust over a reasonably broad range of
parameter values. In contrast,
extreme combinations lead to degenerate segmentations: overly coarse SUs
behave like passage-level chunks, while overly fine SUs lose the
multi-entity binding advantage of hyperedges.

Overall, the sensitivity analysis supports two conclusions. First, the
default setting $(\kappa,d_{\mathrm{eff}})=(75,32)$ achieves the best or
near-best performance. Second, HyperSU is not highly sensitive to moderate
changes in these hyperparameters: a broad range of values still yields
strong performance, while the performance
drops at extreme settings can be explained by clear structural failure
modes in the induced SU segmentation.

\subsection{Hyperedge Statistics and Segmentation Behavior}
\label{app:hyperedge_statistics}

\begin{table}[t]
\centering
\small
\setlength{\tabcolsep}{4pt}
\caption{Statistics of MDL-induced semantic-unit hyperedges on
multi-hop QA datasets. \#SUs counts unique semantic-unit hyperedges.}
\label{tab:su_statistics}
\begin{tabular}{lrrr}
\toprule
\textbf{Dataset} & \textbf{\#Sent.} &
\textbf{\#SUs} & \textbf{Avg. Sent./SU} \\
\midrule
HotpotQA & 58,029 & 18,412 & 3.15 \\
2WikiMultiHopQA & 27,866 & 8,961 & 3.11 \\
MuSiQue & 70,446 & 12,392 & 5.68 \\
\bottomrule
\end{tabular}
\end{table}

Table~\ref{tab:su_statistics} summarizes the segmentation behavior of
the MDL objective. On HotpotQA and 2WikiMultiHopQA, the induced SUs
contain about three sentences on average, indicating that the model does
not degenerate into sentence-level hyperedges. On MuSiQue, the average
SU is longer, which is consistent with the longer and more compositional
reasoning chains in this dataset. Across datasets, MDL segmentation
produces local evidence units rather than passage-level hyperedges,
supporting HyperSU's design goal of constructing reusable,
source-grounded hyperedges from compact contiguous spans.

\subsection{Examples of MDL-Induced Semantic Units}
\label{app:mdl_su_examples}

Table~\ref{tab:mdl_su_examples} shows two semantic units induced from
2WikiMultiHopQA. These examples illustrate that the MDL objective can
keep a coherent event, object, or biographical fact bundle as a single
source-grounded hyperedge while separating it from neighboring topics.

The first example forms a biographical and career-oriented evidence
unit: the person, origin, current location, bands, predecessor band, and
bandmate are all part of the same local context. The second example
forms a production-centered evidence unit: the film, source work,
adapter, director, actors, and roles are tied to the same film record.
Both cases would be awkward to represent only as independent pairwise
edges, because the original group membership and source context are
important for retrieval.

\section{Analysis of Clue-Guided Bidirectional Retrieval}
\label{app:retrieval_analysis}

This section provides a more detailed analysis of the online retrieval stage of
HyperSU. While the main text shows that clue-guided bidirectional retrieval
improves end-to-end performance, here we first isolate the propagation rule
itself and measure SU-level retrieval precision, and then examine how its
forward-exploration and backward-verification hyperparameters affect multi-hop
QA performance. The
results reveal that the two directions play different roles: forward exploration is more sensitive to the reasoning depth of the dataset, whereas backward
verification is most effective when kept shallow and selective.

\begin{table*}[t]
\centering
\small
\setlength{\tabcolsep}{4pt}
\caption{Retrieval-mechanism ablation on the same SU hypergraph. All variants
share the same corpus, entity-aware MDL semantic units, hypergraph, and
SU-to-passage projection, and differ only in how activation propagates over
that structure.}
\label{tab:mechanism_ablation}
\resizebox{\textwidth}{!}{%
\begin{tabular}{@{}lccccccccc@{}}
\toprule
\multirow{2}{*}{\textbf{Propagation}}
& \multicolumn{3}{c}{\textbf{HotpotQA}}
& \multicolumn{3}{c}{\textbf{2WikiMultiHopQA}}
& \multicolumn{3}{c}{\textbf{MuSiQue}} \\
\cmidrule(lr){2-4} \cmidrule(lr){5-7} \cmidrule(lr){8-10}
& R@5 & MAP@5 & GPT-Acc & R@5 & MAP@5 & GPT-Acc & R@5 & MAP@5 & GPT-Acc \\
\midrule
PPR from query entities
& 88.4 & 61.3 & 65.1 & 83.6 & 66.8 & 64.2 & 67.4 & 49.5 & 36.2 \\
PPR from query $\cup$ dense-top-10 entities
& 89.1 & 62.2 & 65.4 & 82.9 & 65.7 & 64.9 & 68.6 & 48.4 & 36.5 \\
Clue-gated expansion, no backward anchoring
& 91.2 & 84.3 & 68.0 & 85.7 & 68.3 & 69.2 & 71.6 & 52.8 & 38.9 \\
\textbf{HyperSU (full)}
& \textbf{95.5} & \textbf{87.8} & \textbf{68.1}
& \textbf{89.1} & \textbf{75.2} & \textbf{72.3}
& \textbf{75.3} & \textbf{59.5} & \textbf{40.2} \\
\bottomrule
\end{tabular}%
}
\end{table*}

\subsection{Retrieval-Mechanism Ablation}
\label{app:mechanism_ablation}

\S\ref{sec:retrieval} defines retrieval as clue-gated bidirectional expansion
over the semantic-unit hypergraph. This subsection isolates that choice. All
variants below share the same corpus, the same entity-aware MDL semantic
units, the same hypergraph, and the same SU-to-passage projection; they differ
only in how activation propagates over that structure. Two variants replace
our propagation with personalized PageRank (PPR), seeded either at
query-linked entities or at query-linked entities together with the entities
of the top-10 dense-retrieved passages, so that PPR receives the same
dense-retrieval signal that candidate-evidence anchoring consumes. A third
variant keeps clue-gated forward expansion but removes backward anchoring and
convergence verification.

\noindent\textbf{Metrics.}
Besides evidence Recall@5 and GPT-Acc., we report MAP@5, which measures how
highly gold passages are ranked inside the reader's top-5 context budget. Let
$p_k$ be the passage ranked at position $k$ and $G$ the set of gold supporting
passages. Then
\[
    \mathrm{AP@5}
    =
    \frac{1}{|G|}\sum_{k=1}^{5}\mathrm{P@}k\cdot\mathbf{1}[p_k\in G],
\]
where $\mathrm{P@}k$ is the precision among the top-$k$ retrieved passages,
and MAP@5 is the macro-average of $\mathrm{AP@5}$ over questions.

\begin{table}[t]
\centering
\footnotesize
\setlength{\tabcolsep}{3pt}
\caption{SU-level retrieval precision with and without backward anchoring. An
SU counts as relevant if it contains a gold supporting sentence;
SU-Prec@$K$ is the proportion of relevant SUs among the top-$K$ ranked SUs,
and the noisy-SU ratio at $K$ is $100-\mathrm{SU\text{-}Prec}@K$.}
\label{tab:su_precision}
\begin{tabular*}{\columnwidth}{@{\extracolsep{\fill}}lccc@{}}
\toprule
\textbf{Variant} &
\textbf{SU-Prec@3} & \textbf{SU-Prec@10} & \textbf{Noisy-SU@3} \\
\midrule
\multicolumn{4}{@{}l}{\textit{HotpotQA}} \\
Full    & \textbf{77.2} & \textbf{37.4} & \textbf{22.8} \\
w/o bwd & 72.5 & 34.8 & 27.5 \\
\cmidrule(lr){1-4}
\multicolumn{4}{@{}l}{\textit{2WikiMultiHopQA}} \\
Full    & \textbf{65.9} & \textbf{31.6} & \textbf{34.1} \\
w/o bwd & 58.0 & 25.5 & 42.0 \\
\cmidrule(lr){1-4}
\multicolumn{4}{@{}l}{\textit{MuSiQue}} \\
Full    & \textbf{51.8} & \textbf{30.7} & \textbf{48.2} \\
w/o bwd & 46.1 & 30.1 & 53.9 \\
\bottomrule
\end{tabular*}
\end{table}

Table~\ref{tab:mechanism_ablation} shows where PPR loses. It still reaches
most gold evidence---on HotpotQA its Recall@5 trails the full model by 7.1
points---but it fails to rank that evidence into the top-5 reader budget,
where MAP@5 trails by 26.5 points. The cause is that the PPR transition
kernel is purely structural: stationary mass grows with entity degree, so
high-degree entities attract activation regardless of query relevance.
Measured directly on HotpotQA, with hubs defined as the top-1\%
highest-degree entities and seed entities excluded, PPR places 37.6\% of its
propagated entity-side mass on hub entities, against 22.9\% under clue-gated
activation, and 53.2\% of its top-5 passages are hub-containing non-gold
passages, against 27.3\% for HyperSU. Among the non-gold passages each method
retrieves, 82\% of PPR's are hub-containing versus 44\% of ours. The second
condition gives PPR the same dense-retrieval information that backward
anchoring consumes, so the remaining gap is not explained by that extra
signal.

Removing backward anchoring degrades ranking more than coverage. On HotpotQA,
GPT-Acc.\ is essentially unchanged, from 68.1 to 68.0, while MAP@5 drops 3.5
points; on 2WikiMultiHopQA and MuSiQue both degrade. This is consistent with
the role of backward anchoring as a precision-oriented refinement rather than
an additional source of recall.

Table~\ref{tab:su_precision} evaluates the same ablation at the SU level.
Removing backward anchoring reduces SU-Prec@3 by 4.7, 7.9, and 5.7 points on
HotpotQA, 2WikiMultiHopQA, and MuSiQue, and the noisy-SU ratio at 3 rises by
the same margins. The drops are smaller at SU-Prec@10 than at SU-Prec@3,
indicating that backward anchoring and convergence-based reweighting mainly
improve the highest-ranked SUs, which are the ones that dominate the capped
mean used for passage scoring.

\noindent\textbf{Bridge-evidence correctness.}
Whether an SU actually bridges two reasoning hops cannot be read off the
existing gold annotations, which mark supporting evidence but not its bridging
role. We therefore evaluate it manually. For each dataset we sample 100 bridge
or compositional questions and inspect every doubly confirmed SU in
$\mathcal{U}_{\mathrm{meet}}$, counting an SU as correct if it comes from a
gold supporting passage and contains an intermediate entity or fact linking
the reasoning hops. Two of the authors independently annotated the sampled
SUs; disagreements were resolved by discussion.
Correctness is 69.7\% on HotpotQA, 77.5\% on
2WikiMultiHopQA, and 68.3\% on MuSiQue, so roughly seven in ten SUs promoted
by convergence verification carry valid bridge evidence.

\begin{table}[t]
\centering
\small
\setlength{\tabcolsep}{4pt}
\caption{Context Relevance on GraphRAG-Bench with and without backward
anchoring.}
\label{tab:ctxrel_backward}
\begin{tabular*}{\columnwidth}{@{\extracolsep{\fill}}lcc@{}}
\toprule
\textbf{Domain--Task} & \textbf{Full} & \textbf{w/o Backward} \\
\midrule
Novel Fact Retrieval        & 78.45 & 74.21 \\
Novel Complex Reasoning     & 67.38 & 61.92 \\
Novel Contextual Summ.      & 88.19 & 82.16 \\
Novel Creative Generation   & 76.34 & 71.89 \\
Medical Fact Retrieval      & 77.52 & 74.13 \\
Medical Complex Reasoning   & 70.14 & 67.65 \\
Medical Contextual Summ.    & 90.04 & 85.17 \\
Medical Creative Generation & 74.93 & 73.58 \\
\midrule
\textbf{Average}            & \textbf{77.87} & 73.84 \\
\bottomrule
\end{tabular*}
\end{table}

On GraphRAG-Bench the corresponding precision-oriented metric is Context
Relevance. Table~\ref{tab:ctxrel_backward} shows that removing backward
anchoring lowers it in all eight domain--task settings, from 77.87 to 73.84
on average. Together with the SU-level precision and manual bridge-evidence
results, this gives direct evidence that backward anchoring and
convergence-based reweighting improve the quality of top-ranked evidence
rather than only its quantity.

\subsection{Sensitivity to Forward-Exploration Hyperparameters}
\label{app:forward_hparam}

Forward exploration starts from query-linked entities and expands through
clue-activated SUs to discover bridge evidence. We analyze two hyperparameters
that control this process: the SU activation threshold $\delta$ and the forward
expansion depth $T^{\mathrm{fwd}}$.

\begin{figure}[t]
    \centering
    \includegraphics[width=\linewidth]{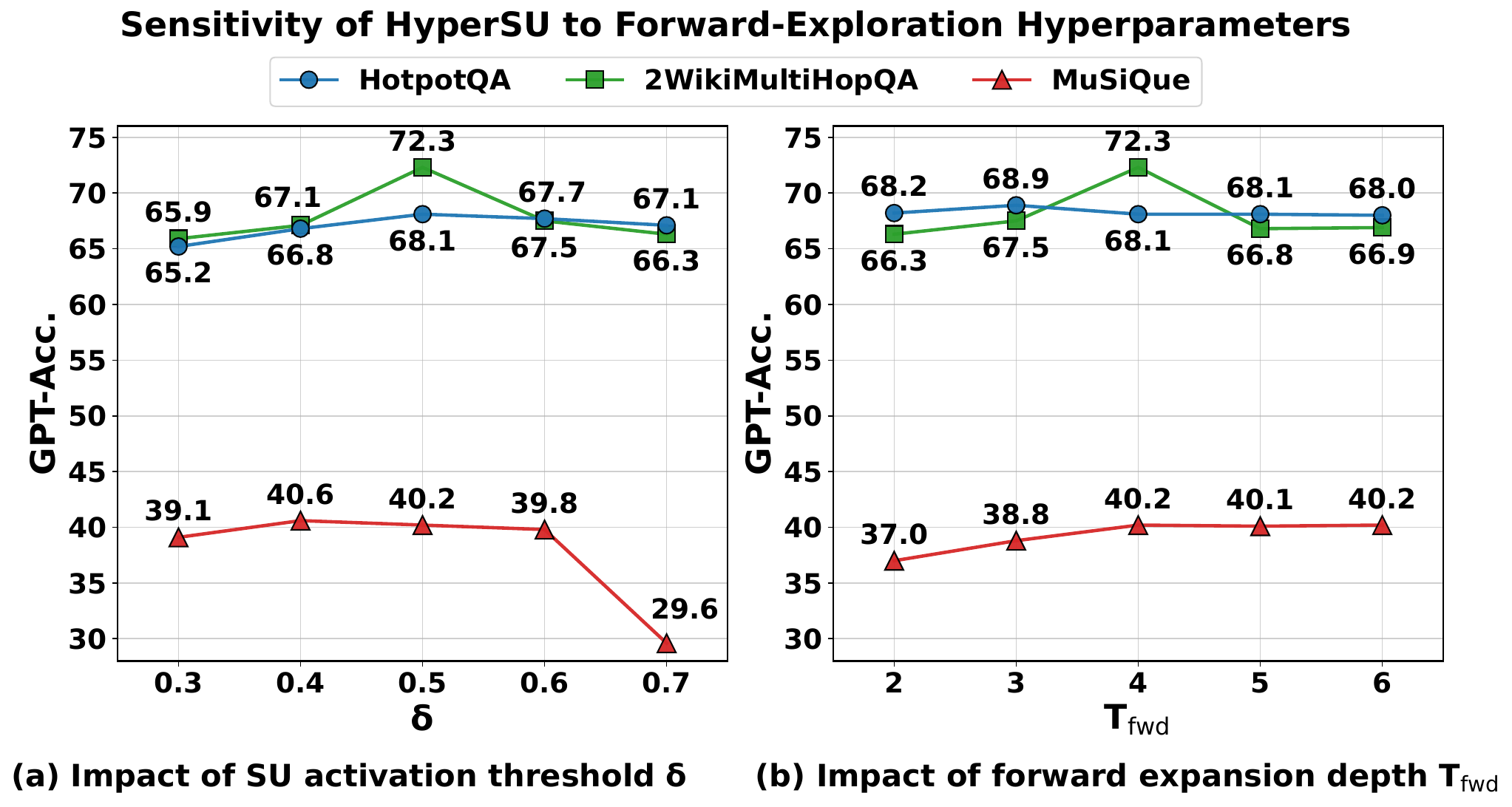}
    \caption{Sensitivity of HyperSU to forward-exploration hyperparameters on HotpotQA, 2WikiMultiHopQA, and MuSiQue. Each point reports GPT-Acc.\ on the corresponding dataset. (a) Impact of the activation threshold $\delta$. (b) Impact of the forward expansion depth $T^{\mathrm{fwd}}$.}
    \label{fig:forward_hparam}
\end{figure}

Figure~\ref{fig:forward_hparam}(a) shows that the effect of $\delta$ depends on
the difficulty of the dataset. On HotpotQA, performance is relatively stable
across different thresholds, ranging from 65.2 to 68.1 GPT-Acc. This suggests
that for shorter reasoning chains, HyperSU is not overly sensitive to the exact
activation threshold as long as extremely noisy SUs are not widely activated.
2WikiMultiHopQA shows a sharper optimum at $\delta=0.5$, where GPT-Acc. reaches
72.3. This indicates that 2WikiMultiHopQA benefits from a balanced threshold:
a lower threshold admits more weakly related SUs, while a higher threshold may
remove useful bridge SUs needed for cross-entity reasoning.

The trend is more pronounced on MuSiQue. Although moderate thresholds produce
similar performance, setting $\delta$ too high causes a large drop, with GPT-Acc.
falling to 29.6 at $\delta=0.7$. MuSiQue contains longer and more difficult
reasoning chains, where some necessary intermediate SUs may be only indirectly
related to the surface form of the query. A strict activation threshold can
therefore filter out semantically indirect but structurally necessary bridge
evidence before expansion can connect it to answer-bearing passages.

Figure~\ref{fig:forward_hparam}(b) further shows that the preferred forward depth
also varies with reasoning complexity. On HotpotQA, performance remains nearly
flat as $T^{\mathrm{fwd}}$ increases, suggesting that shallow or moderate
expansion is already sufficient for most questions. On 2WikiMultiHopQA, the best
performance is obtained at $T^{\mathrm{fwd}}=4$, while deeper expansion degrades
performance. This suggests that moderate expansion helps reach bridge evidence,
but excessive hops can propagate into irrelevant hyperedges and introduce
answer-irrelevant SUs.

MuSiQue exhibits a different pattern. Its performance is lower when
$T^{\mathrm{fwd}}$ is too small, but improves once the depth reaches 4 and then
remains stable. This indicates that harder multi-hop questions
require deeper forward exploration: with too few expansion steps, HyperSU may fail
to collect a complete evidence chain. Overall, these results show that forward
exploration should be sufficiently deep for long-hop reasoning, but bounded to
avoid uncontrolled diffusion into noisy hyperedges.

\subsection{Sensitivity to Verification-Related Hyperparameters}
\label{app:verification_hparam}

\begin{figure}[t]
    \centering
    \includegraphics[width=\linewidth]{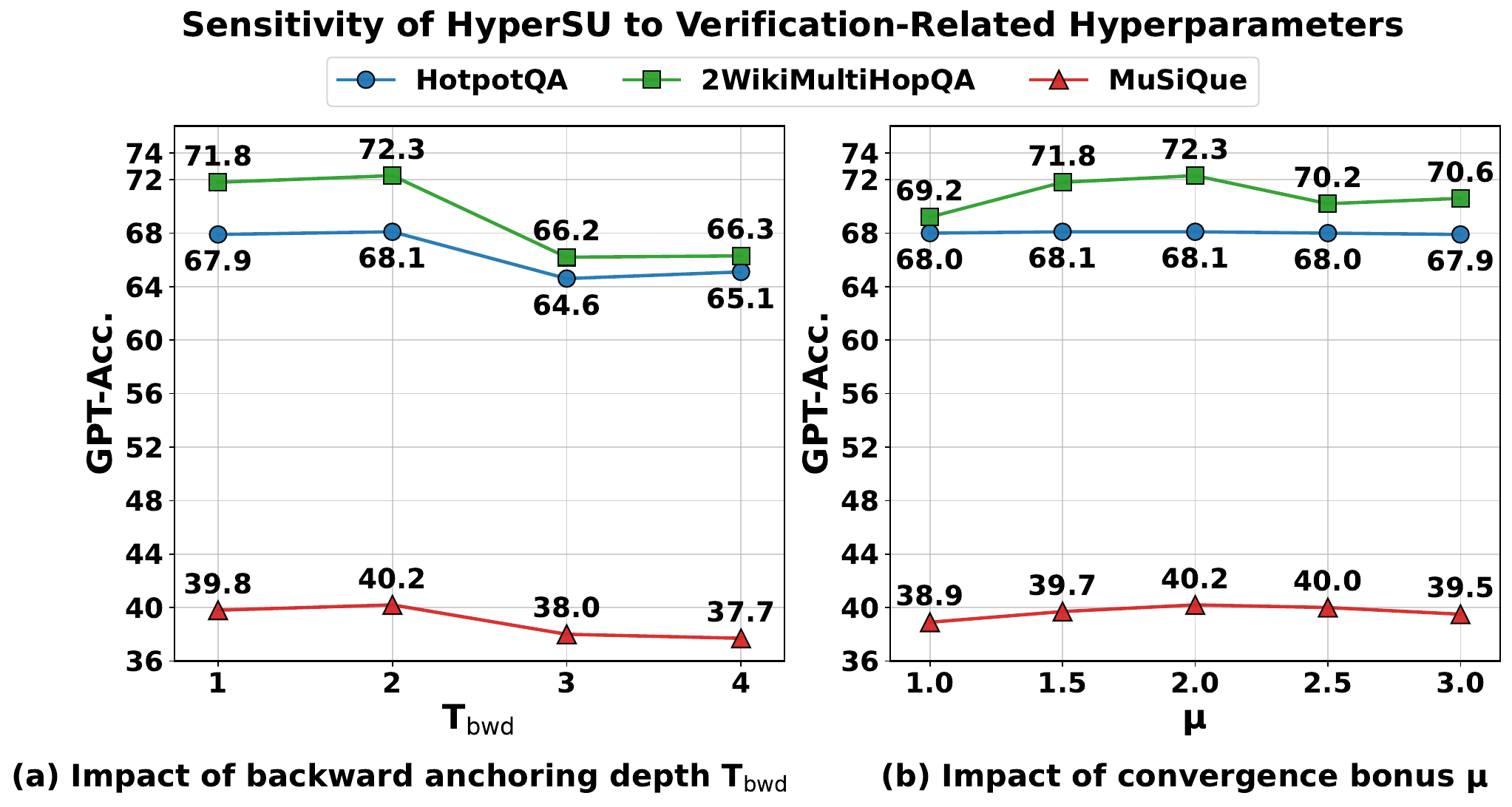}
    \caption{Sensitivity of HyperSU to verification-related hyperparameters on HotpotQA, 2WikiMultiHopQA, and MuSiQue. Each point reports GPT-Acc.\ on the corresponding dataset. (a) Impact of the backward anchoring depth $T^{\mathrm{bwd}}$. (b) Impact of the convergence bonus $\mu$.}
    \label{fig:verification_hparam}
\end{figure}

Backward anchoring starts from entities in top dense-retrieved passages and
verifies forward-reached evidence against those candidates. We analyze two
hyperparameters that control this verification process: the backward anchoring
depth $T^{\mathrm{bwd}}$ and the convergence bonus $\mu$.

Figure~\ref{fig:verification_hparam}(a) shows a consistent trend across all three
datasets: backward verification works best with a small depth. In particular,
$T^{\mathrm{bwd}}=2$ gives the best performance on HotpotQA, 2WikiMultiHopQA, and
MuSiQue. Increasing the depth beyond 2 consistently hurts performance. For
example, on 2WikiMultiHopQA, GPT-Acc. drops from 72.3 at $T^{\mathrm{bwd}}=2$ to
66.2 at $T^{\mathrm{bwd}}=3$; on MuSiQue, it decreases from 40.2 to 38.0. This
suggests that candidate-evidence anchoring should remain local and selective. When
the backward frontier expands too far from dense-retrieved passages, it is more
likely to include distractor entities and topically related but answer-irrelevant
SUs, weakening its ability to verify useful forward evidence.

Figure~\ref{fig:verification_hparam}(b) shows that the convergence bonus is most
effective at a moderate value. Across the three datasets, $\mu=2.0$ achieves the
best or tied-best performance. The gain is especially visible on 2WikiMultiHopQA,
where GPT-Acc. increases from 69.2 at $\mu=1.0$ to 72.3 at $\mu=2.0$, and on
MuSiQue, where it improves from 38.9 to 40.2. This confirms that bidirectional
convergence is a useful verification signal: an SU reached from both the
query-side expansion and the candidate-evidence anchoring frontier is more likely to
serve as reliable bridge evidence.

\begin{table*}[!t]
\centering
\small
\setlength{\tabcolsep}{3pt}
\renewcommand{\arraystretch}{1.08}
\caption{Qualitative case study with top-5 retrieved chunks on a real 2WikiMultiHopQA question. Bold chunk IDs indicate gold supporting chunks: Chunk 577 provides the film--director bridge, and Chunk 578 contains the answer-bearing death-place evidence.}
\label{tab:case_study_top5_chunks}

\newcommand{\traceblock}[1]{%
\par\vspace{2pt}
{\scriptsize
\emph{Top-5 retrieved chunks:}\par
#1
}%
}

\begin{tabular}{@{}
>{\raggedright\arraybackslash}p{0.13\textwidth}
>{\raggedright\arraybackslash}p{0.50\textwidth}
>{\raggedright\arraybackslash}p{0.12\textwidth}
>{\raggedright\arraybackslash}p{0.20\textwidth}
@{}}
\toprule
\textbf{Field / Method} & \textbf{Evidence in Retrieved Context} & \textbf{Outcome} & \textbf{Diagnosis} \\
\midrule
\rowcolor{gray!8}
Question ID & \multicolumn{3}{>{\raggedright\arraybackslash}p{0.82\textwidth}@{}}{352} \\
\rowcolor{gray!8}
Question & \multicolumn{3}{>{\raggedright\arraybackslash}p{0.82\textwidth}@{}}{\textit{Where did the director of film Dancing In The Rain (Film) die?}} \\
\rowcolor{gray!8}
Ground Truth & \multicolumn{3}{>{\raggedright\arraybackslash}p{0.82\textwidth}@{}}{Ljubljana} \\
\rowcolor{gray!8}
Support Context & \multicolumn{3}{>{\raggedright\arraybackslash}p{0.82\textwidth}@{}}{$\surd$ Chunk 577: \textit{Dancing in the Rain} was directed by Bo\v{s}tjan Hladnik. \quad $\surd$ Chunk 578: Hladnik died in Ljubljana in 2006.} \\
\midrule

HippoRAG 2
& Top-5 contains the film chunk and several topical distractors, but misses the Hladnik biography.
\traceblock{
[1] \textbf{577}: Ples v de\v{z}ju is a 1961 Slovene film directed by Bo\v{s}tjan Hladnik. Its international English title is ``Dance in the...''\par
[2] 579: Dancing in the Rain may refer to:\par
[3] 5489: Roman Pola\'nski born 18 August 1933 in Paris; original name Raymond Thierry Liebling is a French-Polish film director, producer...\par
[4] 4931: Barry Levinson born April 6, 1942 is an American filmmaker, screenwriter, and actor...\par
[5] 950: Bernardo Bertolucci 16 March 1941--26 November 2018 was an Italian director and screenwriter...
}
& $\times$ Not specified
& Finds the film page but diffuses through the broad film-director neighborhood. \\

\addlinespace[2pt]

HyperGraphRAG
& Local generated hyperedges can capture the film--director association, but the death-place evidence is in a separate biography chunk.
\traceblock{
[1] \textbf{577}: Ples v de\v{z}ju is a 1961 Slovene film directed by Bo\v{s}tjan Hladnik. Its international English title is ``Dance in the...''\par
[2] 579: Dancing in the Rain may refer to:\par
[3] 950: Bernardo Bertolucci 16 March 1941--26 November 2018 was an Italian director and screenwriter...\par
[4] 4050: Ernst Ingmar Bergman 14 July 1918--30 July 2007 was a Swedish director, writer, and producer...\par
[5] 4931: Barry Levinson born April 6, 1942 is an American filmmaker, screenwriter, and actor...
}
& $\times$ No answer evidence
& Local hyperedges preserve the bridge name but do not reliably compose to the answer-bearing biography. \\

\addlinespace[2pt]

Hyper-RAG
& Diffusion keeps evidence around films and directors, but does not surface Chunk 578.
\traceblock{
[1] \textbf{577}: Ples v de\v{z}ju is a 1961 Slovene film directed by Bo\v{s}tjan Hladnik. Its international English title is ``Dance in the...''\par
[2] 5489: Roman Pola\'nski born 18 August 1933 in Paris; original name Raymond Thierry Liebling is a French-Polish film director, producer...\par
[3] 4931: Barry Levinson born April 6, 1942 is an American filmmaker, screenwriter, and actor...\par
[4] 950: Bernardo Bertolucci 16 March 1941--26 November 2018 was an Italian director and screenwriter...\par
[5] 4050: Ernst Ingmar Bergman 14 July 1918--30 July 2007 was a Swedish director, writer, and producer...
}
& $\times$ No answer evidence
& The signal remains in a topical director region instead of converging on the specific Hladnik page. \\

\addlinespace[2pt]

\rowcolor{gray!10}
\textbf{HyperSU}
& Ranks both supporting chunks in the top context: Chunk 577 identifies Bo\v{s}tjan Hladnik as the director, and Chunk 578 states that Hladnik died in Ljubljana.
\traceblock{
[1] \textbf{577}: Ples v de\v{z}ju is a 1961 Slovene film directed by Bo\v{s}tjan Hladnik. Its international English title is ``Dance in the...''\par
[2] \textbf{578}: Bo\v{s}tjan Hladnik 30 January 1929--30 May 2006 was a Yugoslavian/Slovene filmmaker. Hladnik died in Ljubljana...\par
[3] 579: Dancing in the Rain may refer to:\par
[4] 4931: Barry Levinson born April 6, 1942 is an American filmmaker, screenwriter, and actor...\par
[5] 950: Bernardo Bertolucci 16 March 1941--26 November 2018 was an Italian director and screenwriter...
}
& $\surd$ Ljubljana
& Forward and backward activation meet on the bridge entity and promote both supporting chunks. \\
\bottomrule
\end{tabular}
\end{table*}

\begin{table*}[t]
\centering
\footnotesize
\setlength{\tabcolsep}{3pt}
\renewcommand{\arraystretch}{1.12}
\caption{Qualitative HyperSU activation trace for the retrieval instance in Table~\ref{tab:case_study_top5_chunks}.}
\label{tab:case_activation}
\begin{tabular}{@{}p{0.33\textwidth}p{0.18\textwidth}p{0.14\textwidth}p{0.14\textwidth}p{0.14\textwidth}@{}}
\toprule
\textbf{Activated SU / Evidence} & \textbf{Matched clue} & \textbf{Expansion} & \textbf{Projection} & \textbf{Role} \\
\midrule
\rowcolor{gray!10}
Chunk 577: film title and director Bo\v{s}tjan Hladnik & film, director & forward + backward & Chunk 577 & bridge evidence \\
\rowcolor{gray!10}
Chunk 578: Hladnik biography with ``died in Ljubljana'' & director, death place & forward + backward & Chunk 578 & answer evidence \\
Chunk 577: novel, Paris apprenticeship, DVD materials & partial film context & forward only & lower rank & local distractor \\
Disambiguation and generic film-director biographies & topical overlap & no meeting path & suppressed & distractor \\
\bottomrule
\end{tabular}
\end{table*}

However, larger convergence bonuses do not further improve performance. When
$\mu$ is too large, doubly confirmed SUs can dominate the final passage ranking,
even though convergence alone does not guarantee that an SU is answer-critical.
Some converged SUs may correspond to hub-like entities or broadly relevant
background passages, while some necessary bridge SUs may only be reached from the
forward direction within the limited backward depth. Therefore, $\mu$ should be
treated as a soft preference for bidirectionally verified evidence rather than a
hard preference that suppresses forward-only bridge evidence.

Taken together, these results show that forward exploration is the
main source of retrieval flexibility in HyperSU. Longer-hop datasets
benefit from sufficiently deep expansion and can be more sensitive to
overly strict activation thresholds. Nevertheless, the fixed global
setting used in the main experiments, $\delta=0.5$ and
$T^{\mathrm{fwd}}=4$, provides a balanced operating point across the
evaluated multi-hop QA datasets.

\begin{table*}[t]
\centering
\small
\setlength{\tabcolsep}{5pt}
\caption{Non-LLM evaluation on multi-hop QA benchmarks. We report answer token-level F1 and passage-level evidence Recall@5 (R@5) against gold supporting passages. Best results are in bold and second-best results are underlined.}
\label{tab:non_llm_multihop}
\begin{tabular}{lcccccc}
\toprule
\multirow{2}{*}{Method} & \multicolumn{2}{c}{HotpotQA} & \multicolumn{2}{c}{2WikiMultiHopQA} & \multicolumn{2}{c}{MuSiQue} \\
\cmidrule(lr){2-3} \cmidrule(lr){4-5} \cmidrule(lr){6-7}
& F1 & R@5 & F1 & R@5 & F1 & R@5 \\
\midrule
RAG            & 68.7 & 89.4 & 58.4 & 72.4 & 43.5 & 64.4 \\
HippoRAG 2     & 72.5 & 92.1 & 67.3 & 85.2 & 47.2 & 70.2 \\
LightRAG       & 71.8 & 87.7 & 63.8 & 79.5 & 45.7 & 65.1 \\
LinearRAG      & 69.5 & 90.6 & 65.9 & 84.8 & 44.2 & 69.6 \\
HyperGraphRAG  & \textbf{73.2} & 92.3 & 65.1 & \underline{87.3} & \underline{48.3} & 71.7 \\
Hyper-RAG      & 70.6 & \underline{93.8} & \underline{69.2} & 84.8 & 46.6 & \underline{73.5} \\
HyperSU w/o Clue Agent & 70.3 & 93.2 & 67.4 & 86.1 & 46.7 & 73.0 \\
\textbf{HyperSU} & \underline{72.9} & \textbf{95.5} & \textbf{69.7} & \textbf{89.1} & \textbf{49.0} & \textbf{75.3} \\
\bottomrule
\end{tabular}
\end{table*}

\section{Qualitative Analysis of Retrieval Behavior}
\label{app:qualitative_analysis}

\begin{table*}[t]
\centering
\small
\setlength{\tabcolsep}{4pt}
\caption{Embedding model robustness on GraphRAG-Bench. Each cell reports average ACC over the eight GraphRAG-Bench domain--task settings.}
\label{tab:embedding_model_robustness}
\begin{tabular}{lccccc}
\toprule
\textbf{Method} & \textbf{bge-large-en} & \textbf{e5-large-v2} & \textbf{jina-v3} & \textbf{text-emb-3-large} & \textbf{Avg.} \\
\midrule
RAG            & 54.4 & 54.9 & 55.7 & 56.5 & 55.4 \\
HippoRAG 2     & 60.6 & 61.2 & 60.3 & 62.1 & 61.1 \\
LightRAG       & 54.4 & 54.4 & 54.8 & 55.0 & 54.7 \\
HyperGraphRAG  & 62.6 & 62.5 & \underline{63.2} & \underline{63.9} & 63.1 \\
Hyper-RAG      & \underline{63.2} & \underline{63.7} & 62.9 & \underline{63.9} & \underline{63.4} \\
\textbf{HyperSU} & \textbf{68.7} & \textbf{67.9} & \textbf{68.2} & \textbf{69.0} & \textbf{68.5} \\
\bottomrule
\end{tabular}
\end{table*}

\begin{table*}[t]
\centering
\small
\setlength{\tabcolsep}{5pt}
\caption{LLM backbone robustness on GraphRAG-Bench. For HyperSU, GLiNER-based indexing is fixed; the LLM backbone affects the Clue Agent and reader LLM.}
\label{tab:llm_backbone_robustness}
\begin{tabular}{lcccc}
\toprule
\textbf{Method} & \textbf{GPT-4o-mini} & \textbf{Gemini 2.5 Flash} & \textbf{Claude Sonnet 4.6} & \textbf{Avg.} \\
\midrule
RAG            & 54.4 & 58.7 & 59.9 & 57.7 \\
HippoRAG 2     & 60.6 & 65.4 & 64.0 & 63.3 \\
LightRAG       & 54.4 & 58.3 & 59.1 & 57.3 \\
HyperGraphRAG  & 62.6 & 67.9 & 65.8 & 65.4 \\
Hyper-RAG      & \underline{63.2} & \underline{68.2} & \underline{67.5} & \underline{66.3} \\
\textbf{HyperSU} & \textbf{68.7} & \textbf{72.3} & \textbf{71.7} & \textbf{70.9} \\
\bottomrule
\end{tabular}
\end{table*}

\begin{table*}[!t]
\centering
\small
\setlength{\tabcolsep}{5pt}
\caption{Judge LLM robustness on GraphRAG-Bench. Generated answers are fixed, and only the judge LLM is changed.}
\label{tab:judge_llm_robustness}
\begin{tabular}{lcccc}
\toprule
\textbf{Method} & \textbf{GPT-4o-mini} & \textbf{Gemini 2.5 Pro} & \textbf{Claude Sonnet 4.6} & \textbf{Avg.} \\
\midrule
RAG            & 54.4 & 55.8 & 56.2 & 55.5 \\
HippoRAG 2     & 60.6 & 61.1 & 60.9 & 60.9 \\
LightRAG       & 54.4 & 55.0 & 54.7 & 54.7 \\
HyperGraphRAG  & 62.6 & \underline{63.6} & 62.3 & 62.8 \\
Hyper-RAG      & \underline{63.2} & 63.4 & \underline{63.4} & \underline{63.3} \\
\textbf{HyperSU} & \textbf{68.7} & \textbf{69.9} & \textbf{70.6} & \textbf{69.7} \\
\bottomrule
\end{tabular}
\end{table*}

This section provides a qualitative analysis of how HyperSU retrieves and ranks
multi-hop evidence. We focus on a representative 2WikiMultiHopQA example where
the question requires connecting a film page to the biography page of its
director.

Table~\ref{tab:case_study_top5_chunks} compares HyperSU with representative RAG
baselines. The baselines either retrieve only partial film-side evidence or stay
within a topical neighborhood of films and directors, but fail to surface the
answer-bearing biography chunk. In contrast, HyperSU ranks both supporting chunks
in the final context and produces the correct answer.

Table~\ref{tab:case_activation} further shows the SU-level activation trace for
the same instance. The film-title/director SU and the director-biography SU are
both reached by forward and backward activation, so their projected chunks are
promoted in the final ranking. This example illustrates that HyperSU does not
feed isolated SUs directly to the reader LLM. Instead, clue-activated and
convergence-confirmed SUs are projected back to passages, allowing the final
context to preserve complete chunk-level evidence while still benefiting from
fine-grained SU-level retrieval.

\section{Robustness and Evaluation Reliability}
\label{app:robustness_reliability}

This appendix complements the main evaluation with additional checks on metric choice, model choice, and evaluator choice. The main text reports end-to-end performance using benchmark-specific LLM-based evaluation protocols. While such protocols are standard for open-ended generation tasks, they may introduce sensitivity to evaluator preferences or surface-form variation. We therefore conduct four additional analyses. First, we evaluate multi-hop QA with deterministic answer and evidence metrics. Second, we test whether the conclusions depend on the embedding model or the generation backbone. Third, we fix generated answers and vary only the judge LLM to examine evaluator stability. Fourth, we repeat the comparison against hypergraph-based baselines over multiple runs and test whether the improvements are statistically significant. These analyses are intended to assess whether HyperSU's gains persist under alternative, controlled evaluation settings rather than under a single model or scoring configuration.

\subsection{Non-LLM Evaluation on Multi-hop QA}
\label{app:non_llm_multihop}

To reduce dependence on LLM-as-judge evaluation, we further evaluate multi-hop QA using deterministic metrics. In addition to Contain-Acc.\ and GPT-Acc.\ in the main text, we report answer token-level F1 and passage-level evidence Recall@5. Answer F1 is computed with standard normalization, including lowercasing and punctuation/article removal. Evidence Recall@5 measures whether gold supporting passages appear in the top-5 retrieved passages. These two metrics capture complementary aspects of the task: answer F1 measures whether the final response matches the gold answer at the lexical level, while Recall@5 directly evaluates whether the retriever surfaces the evidence required for multi-hop reasoning.

Table~\ref{tab:non_llm_multihop} shows that HyperSU achieves the highest Recall@5 on all three datasets, indicating that the proposed SU hypergraph retrieval more consistently brings gold supporting evidence into the top retrieved context. This pattern is particularly important for multi-hop QA, where answer generation often fails when any bridge evidence is missing. At the answer level, HyperSU obtains the best F1 on 2WikiMultiHopQA and MuSiQue, and remains within 0.3 points of the best method on HotpotQA. The HotpotQA result suggests that HyperSU does not uniformly maximize lexical overlap under every dataset, but its consistently higher evidence recall provides stronger support for the retrieval-side claim of the paper. Overall, the deterministic metrics are consistent with the main GPT-based results: HyperSU's advantage is not solely an artifact of LLM-as-judge evaluation.

The table also reports two variants relevant to online cost. LinearRAG performs retrieval without any online LLM call. HyperSU can operate in the same regime: disabling the Clue Agent makes the original query embedding the single clue, and the resulting variant still exceeds LinearRAG on all six dataset--metric combinations, by 0.8 to 2.5 F1 and 1.3 to 3.4 Recall@5 points, while the full model with one Clue Agent call is stronger again. HyperSU can therefore be run in either a zero-online-LLM or a multi-clue mode.

\subsection{Robustness to Retrieval and Generation Models}
\label{app:model_robustness}

We next examine whether HyperSU depends on a particular embedding model or LLM backbone. This is important because HyperSU uses embeddings for semantic matching among queries, clues, SUs, entities, and passages, while the online Clue Agent and reader depend on the selected LLM backbone. For these experiments, the evaluation setting is kept fixed within each comparison, and each cell reports average ACC over the eight GraphRAG-Bench domain--task settings. For HyperSU, GLiNER-based indexing remains fixed; thus, changes in the LLM backbone affect the Clue Agent and reader LLM, but not entity extraction or hyperedge construction.

Table~\ref{tab:embedding_model_robustness} shows that HyperSU remains the strongest method under all four embedding models. Its average ACC ranges from 67.9 to 69.0, with an overall average of 68.5, whereas the strongest baseline achieves an average ACC of 63.4. The relative ranking is therefore stable even when the embedding model is changed. This suggests that the benefit of HyperSU is not tied to a single embedding model, but instead comes from the combination of source-grounded SU hyperedges and clue-guided bidirectional expansion. In particular, although the embedding model affects semantic matching quality for all retrieval methods, HyperSU consistently maintains a substantial margin over both graph-based and hypergraph-based baselines.

\begin{table*}[t]
\centering
\small
\setlength{\tabcolsep}{5pt}
\caption{Statistical significance test over 9 independent runs, reported as
mean $\pm$ standard deviation. GraphRAG-Bench columns average ACC over all
eight domain--task settings (All) and over the six settings covering Complex
Reasoning, Contextual Summarization, and Creative Generation (Reason.-Int.).
Multi-hop QA columns use GPT-Acc. Underlined values mark the stronger
reproduced hypergraph baseline, and $\Delta$ is the absolute gain of HyperSU
over it. Test details are given in Appendix~\ref{app:significance_test}.}
\label{tab:significance_hypergraph}
\begin{tabular}{lccccc}
\toprule
\multirow{2}{*}{\textbf{Method}} &
\multicolumn{2}{c}{\textbf{GraphRAG-Bench}} &
\multicolumn{3}{c}{\textbf{Multi-hop QA}} \\
\cmidrule(lr){2-3} \cmidrule(lr){4-6}
& \textbf{All} & \textbf{Reason.-Int.} &
\textbf{HotpotQA} & \textbf{2WikiMHQA} & \textbf{MuSiQue} \\
\midrule
HyperGraphRAG
& $62.6{\pm}0.34$
& $62.5{\pm}0.42$
& $64.9{\pm}0.51$
& $64.1{\pm}0.58$
& $\underline{37.5{\pm}0.72}$ \\

Hyper-RAG
& $\underline{63.2{\pm}0.37}$
& $\underline{63.1{\pm}0.46}$
& $\underline{65.6{\pm}0.47}$
& $\underline{69.1{\pm}0.52}$
& $36.4{\pm}0.76$ \\

\textbf{HyperSU}
& $\mathbf{68.7{\pm}0.29}$
& $\mathbf{69.0{\pm}0.35}$
& $\mathbf{68.1{\pm}0.40}$
& $\mathbf{72.3{\pm}0.45}$
& $\mathbf{40.2{\pm}0.61}$ \\

\midrule
$\Delta$
& $+5.5$
& $+5.8$
& $+2.5$
& $+3.2$
& $+2.7$ \\
\bottomrule
\end{tabular}
\end{table*}

Table~\ref{tab:llm_backbone_robustness} further varies the LLM backbone used for generation. HyperSU obtains the best average ACC under GPT-4o-mini, Gemini 2.5 Flash, and Claude Sonnet 4.6, with an overall average of 70.9. The gains are consistent across backbones, rather than appearing only with one specific reader or planner. This result is important because HyperSU uses an LLM-based Clue Agent at query time; if the gains were mainly due to a particular planner or reader, one would expect the relative advantage to be unstable when replacing the LLM backbone. Instead, the observed stability suggests that the retrieved SU evidence provides useful context across different generation models.

\subsection{Robustness to Judge LLMs}
\label{app:judge_llm_robustness}

Finally, we examine whether the GraphRAG-Bench conclusions are sensitive to the choice of judge LLM. Unlike the previous experiment, generated answers are fixed in this setting, and only the evaluator is changed. This isolates the effect of judge preference from the effect of retrieval or generation quality. A method whose advantage depends on one judge may indicate evaluator-specific bias; a method that remains consistently strong across judges provides more reliable evidence for the reported ranking.

As shown in Table~\ref{tab:judge_llm_robustness}, HyperSU remains the top-performing method under all three judge LLMs. Its average ACC is 69.7, compared with 63.3 for the strongest baseline average. The absolute scores vary moderately across judges, which is expected for LLM-based evaluation, but the method ranking remains stable. This reduces the concern that the main GraphRAG-Bench conclusion is driven by a single evaluator's preference. Together with the deterministic multi-hop QA results in Table~\ref{tab:non_llm_multihop}, these findings provide complementary evidence that HyperSU's improvements are robust under different evaluation protocols, embedding models, generation backbones, and judge LLMs.

\subsection{Statistical Significance Test}
\label{app:significance_test}

We further test whether HyperSU's improvements over
hypergraph-based RAG baselines are statistically significant. We run
HyperSU, HyperGraphRAG, and Hyper-RAG for 9 independent runs under the
same benchmark protocols, and report mean $\pm$ standard deviation.
For each setting, statistical testing compares HyperSU with the stronger hypergraph baseline between HyperGraphRAG and Hyper-RAG.

The runs differ in scope between HyperSU and the baselines. HyperSU builds
its index once and repeats only query-time inference and evaluation. This
discards no variability. HyperSU's indexing pipeline is deterministic.
GLiNER extraction, embedding, dynamic-programming SU induction, and entity
canonicalization return identical output for a fixed corpus, and no LLM is
called during indexing. Rebuilding the index would therefore reproduce it
exactly. The baselines instead repeat the full pipeline, including
LLM-based hyperedge construction, so their runs differ at the index level as
well. All LLM calls use temperature $0$. The residual variation reflects
nondeterminism of the serving API rather than deliberate sampling.

We compare HyperSU with the stronger baseline in each setting using a
two-sided Welch's $t$-test. Welch's test is appropriate because the two
groups have unequal variances by construction. A paired test does not
apply, since runs are not matched across methods. We use
Holm--Bonferroni correction over the five settings at $\alpha=0.05$.

As shown in Table~\ref{tab:significance_hypergraph}, HyperSU consistently
outperforms both reproduced hypergraph baselines, and all corrected
$p$-values fall below $0.001$. The margins are large relative to the
observed spread. The smallest gain, $+2.7$ on MuSiQue, is about four pooled
standard deviations. The largest, $+5.5$ on GraphRAG-Bench All, is about
sixteen. HyperSU also has the lowest standard deviation in all five
settings. This is consistent with the difference in run scope described
above. HyperSU's hyperedges are fixed once the corpus is fixed, whereas the
baselines regenerate their hyperedges with an LLM in every run. The gains
are therefore stable across repeated runs rather than being caused by a
single random run or evaluation setting.

\balance

\end{document}